\documentclass[aps,pra,superscriptaddress,amsmath,amssymb,preprintnumbers,twocolumn,showpacs]{revtex4}
\usepackage{bm}
\usepackage{braket}
\usepackage{graphicx,color}
\newcommand{\NTT}{NTT Basic Research Laboratories, NTT Corporation, 3-1
Morinosato-Wakamiya, Atsugi, Kanagawa, 243-0198, Japan.}
\newcommand{\Riken}{Theoretical Quantum Physics Laboratory, Cluster for Pioneering Research, RIKEN, 351-0198 Wako-shi, Japan}
\newcommand{\Mich}{Department of Physics, The University of Michigan,
Ann Arbor, MI 48109-1040, USA}
\begin{document}


\title{Non-energy-eigenbasis measurements on an ultra-strongly-coupled system interacting with a driven nonlinear resonator}


\author{Suguru Endo}
\affiliation{\Riken}
\author{Yuichiro Matsuzaki}
\affiliation{\NTT}
\author{Kosuke Kakuyanagi}
\affiliation{\NTT}
\author{Shiro Saito}
\affiliation{\NTT}
 \author{Neill Lambert}
\affiliation{\Riken}
\author{Franco Nori}
\affiliation{\Riken}
\affiliation{\Mich}


\begin{abstract}

 We explore the problem of projecting the
 ground state of a system into a superposition between energy eigenstates when the coupling between measurement device and system is much smaller than the energy scales of the system itself. As
 a specific example, we investigate an ultra-strongly coupled
 light-matter system whose ground state exhibits non-trivial entanglement between the
  atom and photons.
 As a measurement apparatus we consider both linear and
 non-linear driven resonators. We find that the state of the non-linear resonator
 can exhibit a much stronger correlation with the ultra-strongly coupled system than
 the linear resonator, even when the system-measurement apparatus coupling strength is  weak. Also, we investigate the conditions for when the nonlinear resonator can
 be entangled with the ultra-strongly coupled system, which allows us to project the ground state of the
 ultra-strongly coupled system into
a non-energy eigenstate. Our proposal paves the way to realize projective
 measurements in an arbitrary basis,  which would significantly broaden
 the possibilities of controlling quantum devices.

\end{abstract}
\maketitle

A quantum measurement typically projects the a system into an eigenstate of
the measured observable $\hat{A}$.
In quantum measurement theory, the measurement apparatus
interacts with the target system due to an interaction Hamiltonian
$H_I=J\hat{A}\otimes \hat{B}$, where $\hat{B}$ denotes the operator of the
 apparatus and $J$ denotes a coupling strength \cite{measurement1,qndcondition}.  This process induces a correlation
between the system and apparatus.  A subsequent measurement on the
apparatus itself implements the projection of the target system, and
the readout of the apparatus is associated
with the eigenvalues of the system observable $\hat{A}$. To realize a quantum
non-demolition measurement, the observable $\hat{A}$ should commute with
the target system Hamiltonian \cite{qndcondition}. In addition, in several experiments \cite{measurement3,jba2,jba3,jba4},
projective measurements on quantum systems have been demonstrated directly in
the energy-eigenbasis itself, where the
observable $\hat{A}$ is the Hamiltonian of the target system.

Although projective measurements in an arbitrary
  basis would significantly broaden the possibilities of controlling
 quantum states if realized \cite{application1,application2,application3}, such non-energy-eigenbasis measurements are, surprisingly, sometimes not straightforward.
 Ideally, if the system observable $\hat{A}$ to be
  measured does not commute with the system Hamiltonian, $\hat{A}$ has a
  matrix component to induce transitions between the energy
  eigenstates. Importantly, however, if the coupling strength $J$ is much smaller
  than the energy of the system, such transition matrix components
  disappear under a rotating wave approximation \cite{introductory} (see Appendix A for details), and we cannot
  project the system into the eigenbasis of $\hat{A}$; the system stays in its energy eigenbasis.

  On the other hand, if
  the coupling between the system and apparatus is much larger than the
  system energy, one can perform a projective measurement much faster
  than the typical time scale of the system, hence realizing
  non-energy eigenbasis measurements~\cite{ashhab,ashhab2,ashhab3}. However, if energy scales are comparable, the dynamics, and the subsequent quantum measurement process,
  becomes much more complicated than the cases described above.
  Understanding the interaction between the apparatus and system, and their dynamics, is important not only
  for explaining the mechanism of quantum
  projective measurements but also to achieve a higher level of control over
 quantum states.

  The ultra-strong and deep-strong coupling regimes between atoms and light is an especially
attractive area to explore the possibility of  non-energy eigenbasis measurements. This is
because the ground state of this system exhibits non-trivial entanglement between the
  atom and photons, and virtual excitations, which are difficult to probe with energy eigenbasis measurements alone.
This hybridization of light and matter is one of the core topics in
quantum physics
\cite{hybrid1,hybrid2,hybrid3,bulute11,georgescu12,hybrid4,hybrid5, XinWang,ultra12}.
In addition, when the coupling strength between light and matter becomes extremely
strong, so that it surpasses the cavity resonance frequency, it is
predicted that a new ground state will emerge
\cite{ultra1,ultra2,ultra3,ultra4,protected,roberto2017,transition, felicett,u8,u9,u10, u11, u12, u13, u14, u15, u16, u17}.
Such a regime was recently experimentally demonstrated \cite{expultra1,expultra2,expultra3}.

Measurements of ultra-strongly-coupled systems so far have mostly focused on extraction of photons from the ground state, via modulation of some system parameter \cite{Simone, christian} (akin to approaches used to observe the dynamical Casimir effect \cite{dcasimir3,dcasimir2,dcasimir1,dcasimir4,dcasimir5}),
or transitions out of the ground-state itself \cite{Roberto, Mauro_dqd}. \textcolor{black}{In addition, another proposal suggested using an ancillary qubit coupled to an ultra-strongly coupled system \cite{Ciuti}, with the goal of doing QND measurement of the photons in the ground state. }
Since the ground state of the ultra-strongly-coupled
system contains virtual photons, we can also in principle extract the virtual
photons if the state is projected into
a non-energy eigenbasis state \cite{Simone} (akin to approaches used to
observe the dynamical Casimir effect
\cite{dcasimir3,dcasimir2,dcasimir1,dcasimir4,dcasimir5}).

If we could observe such photons extracted from the
ground state, this would be a direct evidence of the implementation of
the non-energy eigenbasis states.
Moreover,
non-eigenbasis measurements on a ground
state of a ultra-strongly-coupled system
could potentially be used to induce an optical cat state, which is itself a resource for quantum
information processing \cite{protected,roberto2017}.
Given these potential benefits, and open problems to be solved,
the ultra-strongly-coupled system is attractive as an example with which to
investigate the problem of non-energy eigenbasis measurements.
 Although there are several previous works studying the quantum
properties of the ground state in an ultra-strongly-coupled system
\cite{Ciuti, Mauro,ultra1,ultra2,ultra3,ultra4}, here we focus only on how
to perform non-energy eigenbasis measurements on the ground state of
such a system.


In this paper, we specifically analyze the dynamics of an ultra-strongly
coupled system interacting with a measurement apparatus,
when the measured system observable does not commute with the
system Hamiltonian. We evaluate the dynamics of the
measurement apparatus during the interaction, the back-action of the
measurements on the system, and the correlations between the system and the
apparatus. Such properties are typically studied when one tries to examine in detail a quantum
measurement process \cite{nakano,trajectory,backaction, continuous}.
Although there exist theoretical proposals to use a detector that continuously monitors the system \cite{ashhab2}, here we consider a binary-outcome measurement
performed on the measurement apparatus {\em after} the measurement apparatus and system have been allowed to interact.
 \textcolor{black}{Such a binary-outcome measurement} is understood to
induce a strong correlation with the system
\cite{jba6,kakuyanagijps},
\textcolor{black}{which is crucial to realize our goal of non-energy eigenbasis
measurements}.

While linear resonators are used as a standard method for quantum
measurement in cavity quantum electrodynamics
and circuit quantum electrodynamics, in some cases a nonlinearity has been employed to improve qubit readout~\cite{jba1,jba2,jba3,jba4,jba5,jba6}.
Due to the bifurcation effect, the state of the nonlinear resonator becomes
highly sensitive to the state of the system, which enables one to implement a high-visibility readout.
Here, with both full numerical modeling and a low-energy approximation, we investigate how such a driven nonlinear resonator
 interacts with the ultra-strongly-coupled system.
Surprisingly, although the coupling between the nonlinear measurement
device and the ultra-strongly-coupled system is weak compared to system energy scales, 
we show that the dynamic evolution can induce a strong correlation between
them, \textcolor{black}{which shows that the non-linear resonator would be a suitable device to realize the non-energy eigenbasis measurements}.
Moreover, we evaluate how much \textcolor{black}{quantum
correlations such as} entanglement
\textcolor{black}{and quantum discord} are
generated between system and measurement device during this
evolution. \textcolor{black}{These results let us know the conditions
when the non-energy eigenbasis measurements can be realized with this system.}

The remainder of this paper is organized as follows.
First, we introduce the ultra-strongly-coupled system and its ground
state. Second, we discuss the interaction between the nonlinear
resonator and the ultra-strongly-coupled system, and we introduce a
coarse-graining measurement of the nonlinear resonator itself.
Third, we present numerical results to show how a strong correlation arises, even
in a parameter regime where the coupling strength may be incorrectly considered to be
negligible. 
\textcolor{black}{Fourth, we show that, as the effective energy of the ultra-strongly-coupled system decreases, the entanglement between the ultra-strongly
coupled system and the nonlinear resonator-increases.} \textcolor{black}{Finally, we examine the quantum discord  between the ultra-strongly
coupled system and the nonlinear resonator.}

\section{Ultra-strong coupling between light and matter}
The Hamiltonian of light in a single-mode cavity ultra-strongly-coupled to matter (where the matter is well described by a two-level system) is, in its simplest form, given by the Rabi model \cite{rabi}
\begin{align}
\hat{H}_{\rm{Rabi}}=\frac{\omega_{\rm{q}}}{2}\hat{\sigma}_x+g (\hat{a}+\hat{a}^\dag)\hat{\sigma}_z+\omega_{\rm{r}} \hat{a}^\dag \hat{a},
\end{align}
where $\hat{a}$ ($\hat{a}^\dag$) is an annihilation (creation)
operator for the single-mode cavity/resonator,  $\omega_{\rm{q}}$ ($\omega_{\rm{r}}$) denotes the qubit
(resonator) frequency, and $g$ is the coupling strength between
resonator (light) and qubit (matter).

Recall that, when the matter is in the form of a superconducting flux qubit, as
in the recent ultra-strong coupling experiments in \cite{expultra1, expultra2, expultra3},
$\hat{\sigma }_z=|L\rangle \langle L|-|R\rangle \langle R|$ is
diagonal in the persistent-current basis of $L$ and $R$ of the superconducting flux
qubit.

\begin{figure}[t]
\centering
\includegraphics[width=\columnwidth]{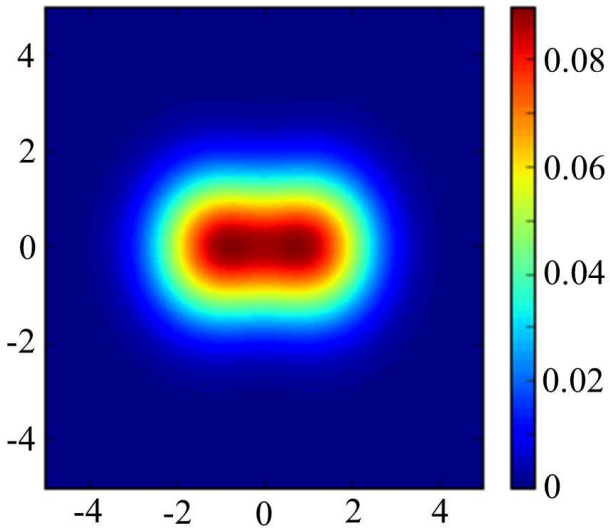}
\caption{(Color online) The $Q$ function $\bra{\beta}\hat{\rho}\ket{\beta}/\pi$ of the reduced density matrix $\hat{\rho}$ of the cavity in the ground state $\ket{G}$. Here $\omega_{\rm{q}}=2\pi \times 0.299$ GHz, $g=2\pi \times 4.920$ GHz, $\omega_{\rm{r}}=2\pi \times 6.336$ GHz.   }
\label{qfun}
\end{figure}

Throughout this paper we assume that the qubit frequency is much
smaller than the resonator frequency, allowing us later to use an
adiabatic approximation. In this case, in the limit $\omega_{\rm{q}} \rightarrow 0$, we can approximately write the ground state
of this system as \cite{ultra1}
\begin{align}
\ket{G}\simeq \frac{1}{\sqrt{2}}(\ket{R}\ket{\alpha}-\ket{L}\ket{-\alpha})
\end{align}
where
\begin{equation}
\alpha=g/\omega_{\rm{r}},
\end{equation}
is the ratio of the coupling strength and
resonator energy.
As an example, using parameters close to those used in \cite{expultra1},  we plot the $Q$ function of the reduced density matrix of $\ket{G}$
where the atom is traced out
in
Fig.\ref{qfun}. \textcolor{black}{The definition of the $Q$ function for a state $\hat{\rho}$ is $Q (\beta)= \frac{1}{\pi} \bra{\beta}\hat{\rho}\ket{\beta}$, where $\ket{\beta}$ is a coherent state for a complex number $\beta$. We plot the real part of the $\beta$ in the $x$ axis while we plot the imaginary part of the $\beta$ in the $y$ axis.}
It is worth mentioning that, if we can realize a projective
measurement in the basis of $\frac{1}{\sqrt{2}}(\ket{R} +\ket{L})$ or
$\frac{1}{\sqrt{2}}(\ket{R} -\ket{L})$ on this ground state, we can
create an optical cat state, $\frac{1}{\sqrt{2}}(\ket{\alpha}
+\ket{-\alpha})$,  in the cavity.

\par

\section{Using a nonlinear resonator as a measurement device}
Here,
 as a measurement apparatus,
we consider a driven nonlinear resonator dispersively
coupled to the qubit.
It is
well understood that a nonlinear resonator can exhibit a bistability \cite{rigo,nakano, clerk1, clerk2,backaction},
which makes such a device sensitive to small changes in external fields.
In addition, the nonlinearity induces a rapid change in the photon
 number under driving  \cite{nakano}, compared to the linear case. 
 When used as a measurement device, the fast evolution and the sensitivity of the steady-state to weak fields results in a strong and fast
 correlation of the nonlinear resonator state with the qubit being measured, potentially giving a means to implement a rapid projective measurement.
One should note that typically the state of
the nonlinear resonator is itself measured by standard homodyne
techniques \cite{rabi}, and this measurement provides
the information about the qubit state.

It is worth mentioning that there are some theoretical proposals to treat such a measurement device as a two level system when the measurement outcomes are binary \cite{ashhab}. However, since such a simplification cannot quantify the strength of the correlation between the target qubit and measurement apparatus during the measurement process, we need to model the measurement apparatus with a proper Hamiltonian as we will describe below.

\par
The total system, composed of the ultra-strongly-coupled light-matter system, and the nonlinear resonator measurement device, can be described by the Hamiltonian \textcolor{black}{in the laboratory frame}~\cite{ultra1,ultra3, ultra4, nakano, clerk1, clerk2, backaction}
\begin{align}
\textcolor{black}{\hat{H}_{\rm{tot}}^{(\rm{lab})}}&\textcolor{black}{=\hat{H}_{\rm{Rabi}}+\hat{H}_{\rm{nr}}^{(\rm{lab})} +\hat{H}_{\rm{int}}^{(\rm{lab})}\label{total_LF}} \\
\textcolor{black}{\hat{H}_{\rm{nr}}^{(\rm{lab})}}&\textcolor{black}{= (\delta+\omega_d) \hat{b} ^\dag \hat{b} - \chi (\hat{b}^\dag \hat{b} )^2 -f \mathrm{cos}~(\omega_d t) ~(\hat{b}+\hat{b}^\dag)} \\
\textcolor{black}{\hat{H}_{\rm{int}}^{(\rm{lab})}}&\textcolor{black}{= J \hat{\sigma}_z \hat{b}^\dag \hat{b}}
\end{align}
where $\hat{b}$ is an annihilation operator of the nonlinear
system, $\delta$ denotes the detuning between the
nonlinear resonator energy and driving frequency, $\chi$ is the nonlinearity strength, $f$
denotes the driving strength of the nonlinear resonator, \textcolor{black}{and $\omega_d$ is the driving frequency of the nonlinear resonator.} In addition, $J$ is the
coupling  between the qubit
and the nonlinear resonator, which is not derived from the dispersive approximation to a dipole coupling, but is intrinsic (\textcolor{black}{see Appendix B for details}.)  \textcolor{black}{In the rotating frame defined by $\hat{U}_{\mathrm{rot}}(t)= \mathrm{exp}[-i\omega_d ~t \hat{b}^\dag \hat{b}]$ and by applying the rotating wave approximation, we have}
\begin{align}
\hat{H}_{\rm{tot}}&=\hat{H}_{\rm{Rabi}}+\hat{H}_{\rm{nr}} +\hat{H}_{\rm{int}}\label{total} \\
\textcolor{black}{\hat{H}_{\rm{nr}}}&\textcolor{black}{= \delta \hat{b} ^\dag \hat{b} - \chi (\hat{b}^\dag \hat{b} )^2 -\frac{f}{2} (\hat{b}+\hat{b}^\dag)} \\
\hat{H}_{\rm{int}}&= J \hat{\sigma}_z \hat{b}^\dag \hat{b},
\end{align}
In order to include the loss of photons from the
nonlinear resonator, we adopt the following Lindblad master equation, valid when the coupling between nonlinear resonator and its environment is weak, and when the coupling $J$ between nonlinear resonator and qubit is weak \cite{nakano, clerk1,clerk2, backaction}
\par
 \begin{align}
 \frac{d}{dt} \hat{\rho} = -i[\hat{H}_{\rm{tot}}, \hat{\rho}]+\frac{\kappa}{2}(2 b\hat{\rho} b^\dag -\hat{b}^\dag \hat{b} \hat{\rho}-\hat{\rho} \hat{b}^\dag \hat{b} ),\label{lind}
 \end{align}
where $\kappa$ denotes the photon leakage rate from the nonlinear cavity.  The potential losses from the ultra-strongly coupling system are described later.

\subsection{Coarse-graining of the measurement outcome}
After the qubit and the measurement apparatus have interacted for some time, we need to implement a measurement
on the measurement apparatus itself. Ideally, one could apply a
projection operator \hspace{1mm}$\hat{P}_x=\ket{x} \bra{x}$ on the nonlinear resonator, where $\ket{x}$ is an eigenvector
of the quadrature operator  \hspace{1mm}$\hat{x}=(\hat{b}+\hat{b}^{\dagger})/2$. However, due to
imperfections in the measurement setup, one cannot resolve arbitrarily
small differences in the state of the resonator.  Normally, to describe more realistically the measurement process, one
takes this into account by considering the integrated signal-to-noise~\cite{bathinte}, where
the noise can include contributions from vacuum fluctuations and noise in the measurement apparatus itself.  Here, instead
 we employ a ``coarse
 graining'' approximation
described by the following operator
\begin{align}
\hat{E}_x= \frac{1}{\pi ^{1/4} \sqrt{2\sigma}}\int _{-\infty}^{\infty} dx^\prime \exp\left[-\frac{(x^\prime-x)^2}{4\sigma^2}\right]\ket{x^\prime}\bra{x^\prime},
\end{align}
where $\sigma$ is the width of the error of the measurement process, and
the post-measurement state is described by $\hat{E}_x\hat{\rho} \hat{E}_x/{\rm{Tr}}[\hat{E}_x\hat{\rho} \hat{E}_x]$. Similar coarse graining approaches have been made in Refs.~\cite{coarse1, coarse2}. This approach allows us to consider
the transition from small to large noise situations without being specific about the source of the noise.

Correlations between the nonlinear resonator and
the qubit should occur after they have interacted for some time, and,  for the parameter regime we use in this work,
 typically the nonlinear resonator
state with $x \geq 0$ ($x <0$) corresponds to an outcome where the qubit was initially in its excited
(ground) state. We can describe the
 post measurement state of the ultra-strongly-coupled (USC) system as \textcolor{black}{(see Appendix C for details)}
\begin{align}
\hat{\rho}_{x \geq 0} &= \frac{1}{N} \int_{-\infty}^{\infty} dx \hspace{1mm}\mathrm{erfc} \bigg(-\frac{x}{\sqrt{2}\sigma} \bigg)\bra{x}\hat{\rho}\ket{x} \\
\hat{\rho}_{x < 0} &= \frac{1}{N^\prime} \int_{-\infty}^{\infty} dx \hspace{1mm}\mathrm{erfc} \bigg(\frac{x}{\sqrt{2}\sigma} \bigg)\bra{x}\hat{\rho}\ket{x},
\end{align}
where $\mathrm{erfc}$ is the complementary error function and $N$ and $N^\prime$ are normalization factors.
\par

In the limit when $\sigma \rightarrow +\infty$, we
obtain $\hat{\rho}_{x \geq 0}=\hat{\rho}_{x < 0} \propto \int _{-\infty}^{\infty} dx  \bra{x}
\hat{\rho} \ket{x}$. In this case,  the measurement results do not contain
any information of the post-measurement state of the qubit. On the other
hand, we obtain
\begin{equation}
\hat{\rho}_{x \geq 0} \propto \int _{0}^{\infty} dx  \bra{x}
\hat{\rho} \ket{x},
\end{equation}
and
\begin{equation}
\hat{\rho}_{x < 0} \propto \int _{-\infty}^{0} dx  \bra{x} \hat{\rho} \ket{x},
\end{equation}
in the limit $\sigma \rightarrow 0$, which corresponds to an ideal
projective measurement that can perfectly distinguish $x\geq 0$ or $x<0$.

\par

\subsection{Losses in ultra-strongly coupled systems}

We must also consider the coupling
between the cavity component of the ultra-strongly-coupled system and its environment.  This
allows us to evaluate properties of the photons leaking out from the system after the measurement on the nonlinear resonator. The
interaction Hamiltonian between the system and the environment can be described as \cite{bathinte}  \begin{equation}\hat{H}_{I}=\int d\omega
\textcolor{black}{\Gamma(\omega)}(\hat{a}+\hat{a}^\dag)(\hat{c}(\omega)+\hat{c}(\omega)^\dag), \end{equation} where $\hat{c}(\omega)$
denotes a boson annihilation operator for the environment (e.g., an open transmission line). When we
move to the Heisenberg picture, defined by the Hamiltonian $\hat{H}_{\rm{Rabi}}$,
the operator $(\hat{a}+\hat{a}^\dag )$ becomes time dependent.
We can  define $(\hat{a}+\hat{a}^\dag )(t)=\hat{X}^++\hat{X}^-$, where
 $\hat{X}^+$ ($\hat{X}^-$) denotes the positive (negative) frequency component~\cite{xoperator1,xoperator2,xoperator3}. With a rotating wave approximation, we have
 \begin{equation}
\hat{H}_{I} \approx
\int d\omega \textcolor{black}{\Gamma(\omega)}[\hat{X}^+ \hat{c}(\omega)^\dag+\hat{X}^-\hat{c}(\omega)].\end{equation}
By using a
Markov approximation $\Gamma(\omega)=\sqrt{\gamma/2\pi}$ with
the standard input-output formalism \cite{bathinte}, we obtain
\begin{align}
\hat{c}_{\mathrm{out}}&= \hat{c}_{\mathrm{in}}+\sqrt{\gamma} \textcolor{black}{\hat{X}^+} \\
\hat{c}_{\mathrm{in}}&\equiv \frac{1}{\sqrt{2\pi}}\int d\omega e^{-i \omega t} \hat{c}(\omega),
\end{align}
where $\hat{c}_{\mathrm{out}}$ ($\hat{c}_{\mathrm{in}}$) denotes an output (input)
 operator. This means that the photons leaking from the USC system into the transmission line are described by the
 operators $\hat{X}^+$ and $\hat{X}^-$.
 We use these definitions to describe the potentially observable real photons in the USC system.  However,
 in the simulations we perform, we assume that the decay time of the USC system is
 longer than all other time scales that we consider, and so we only explicitly take into account the
 decay of the nonlinear measurement apparatus.

\section{Full dynamics of the USC system and nonlinear measurement device}

\begin{figure}[t]
\centering
\includegraphics[width=\columnwidth]{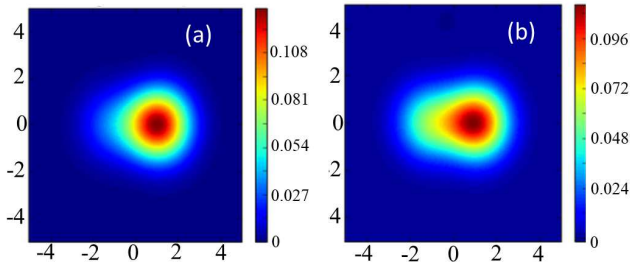}
\caption{(Color online) \hspace{1mm}The $Q$ function of the reduced
 density matrix of the resonator ultra-strongly-coupled to the qubit
 after the coarse graining-measurement. We
 consider in (a) a
projection into $x< 0 $ at a time $t=500 \hspace{1mm} \mathrm{ns}$
for $\sigma=0.5$, and
(b) the same projection for $\sigma=50$.
 These examples confirm that, as we increase the value of $\sigma $,
 the change of the $Q$ function induced by the measurement becomes smaller.
   We set $t=500 \hspace{1mm} \mathrm{ns}$,
 $\omega_{\rm{q}}=2\pi \times 0.299$ $\mathrm{GHz}$, $g=2\pi \times
 4.920\hspace{1mm}\mathrm{GHz}$, $\omega_{\rm{r}}=2\pi \times
 6.336\hspace{1mm}\mathrm{GHz}$, $\kappa=2\pi \times 2.375
 \hspace{1mm}\mathrm{MHz},$ $\delta=2\pi \times 5.698\hspace{1mm}
 \mathrm{MHz}$, $\chi=2\pi\times 80.735 \hspace{1mm}\mathrm{kHz}$,
 $f=2\pi\times 22.792 \hspace{1mm}\mathrm{MHz}$, and $J=2\pi\times
 949.8\hspace{1mm}\mathrm {kHz}$.}
\label{kekka1}
\end{figure}

Using the parameters from \cite{expultra1}, we numerically~\cite{qutip,qutip2} solve
Eq.~(\ref{lind}), with the USC system in the initial state $\ket{G}$, and we consider the post-measurement state of the USC system after the coarse-graining measurement on the nonlinear resonator is performed at the time $t=500 \hspace{1mm}\mathrm{ns}$.
In Fig.~\ref {kekka1}, we show how the $Q$ function of the resonator part
of the USC system depends on the coarse-graining
value.
 One immediately sees a change in the state of the USC resonator when the measurement becomes
 weaker (corresponding to an increase of the coarse-graining values of $\sigma$).

\par

\begin{figure}[t]
\centering
\includegraphics[width=8.5cm]{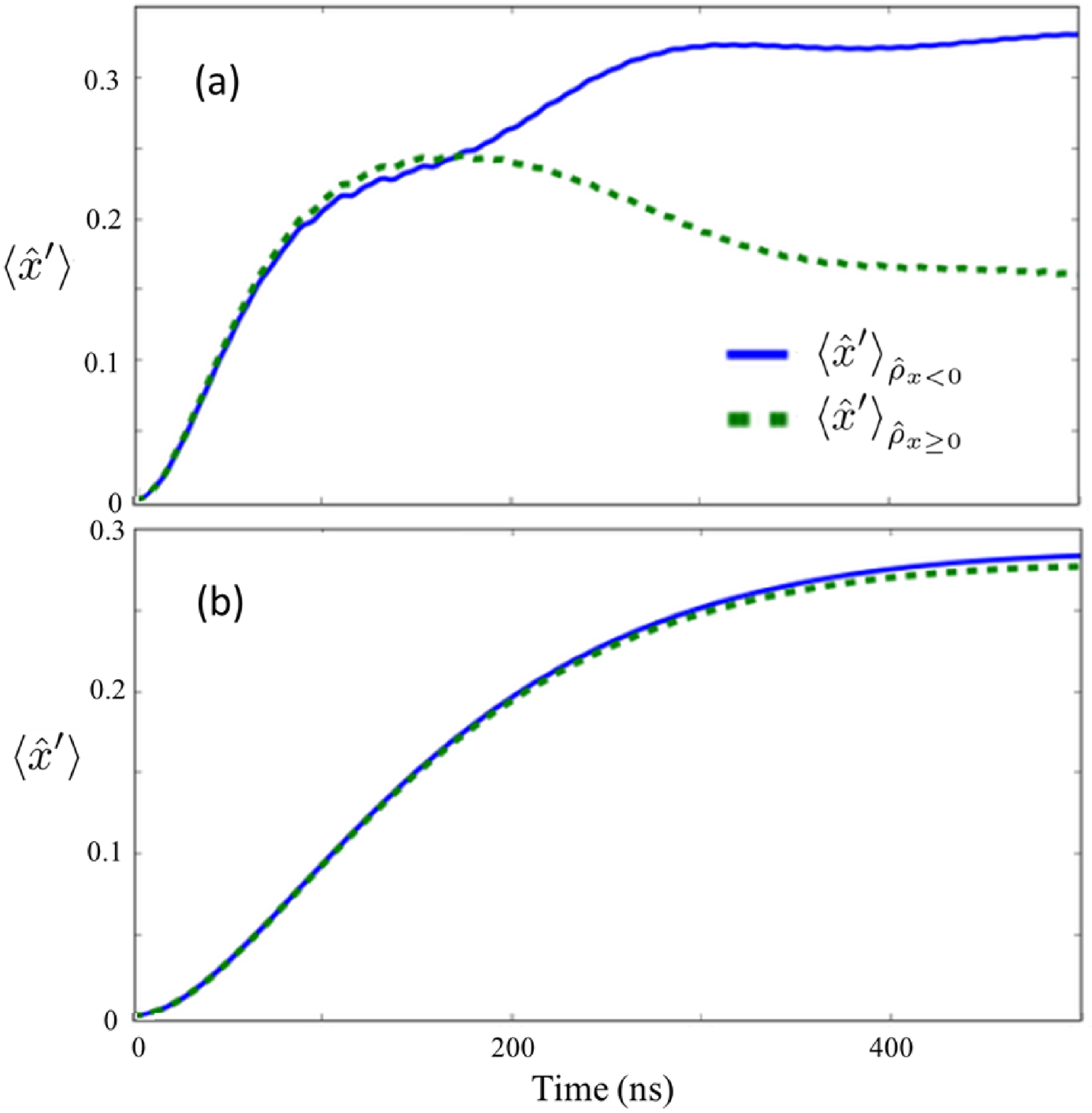}
\caption{(Color online) $\braket{\hat{x}^\prime}$, the quadrature of the cavity in the USC system, after the coarse-graining measurement that projects the
 state into $\hat{\rho}_{x\geq0}$ or $\hat{\rho}_{x<0}$ depending on the measurement
 results. We plot
 $\braket{\hat{x}^\prime}$  in
 (a) for the nonlinear resonator
 and (b) for the linear resonator as the measurement apparatus.
 Here, we set the coarse-graining value as $\sigma=5$. For the other parameters, we use the
 same as those in
 Fig.~\ref{kekka1}.}
\label{kekka2}
\end{figure}

\begin{figure}[t]
\centering
\includegraphics[width=8.5cm]{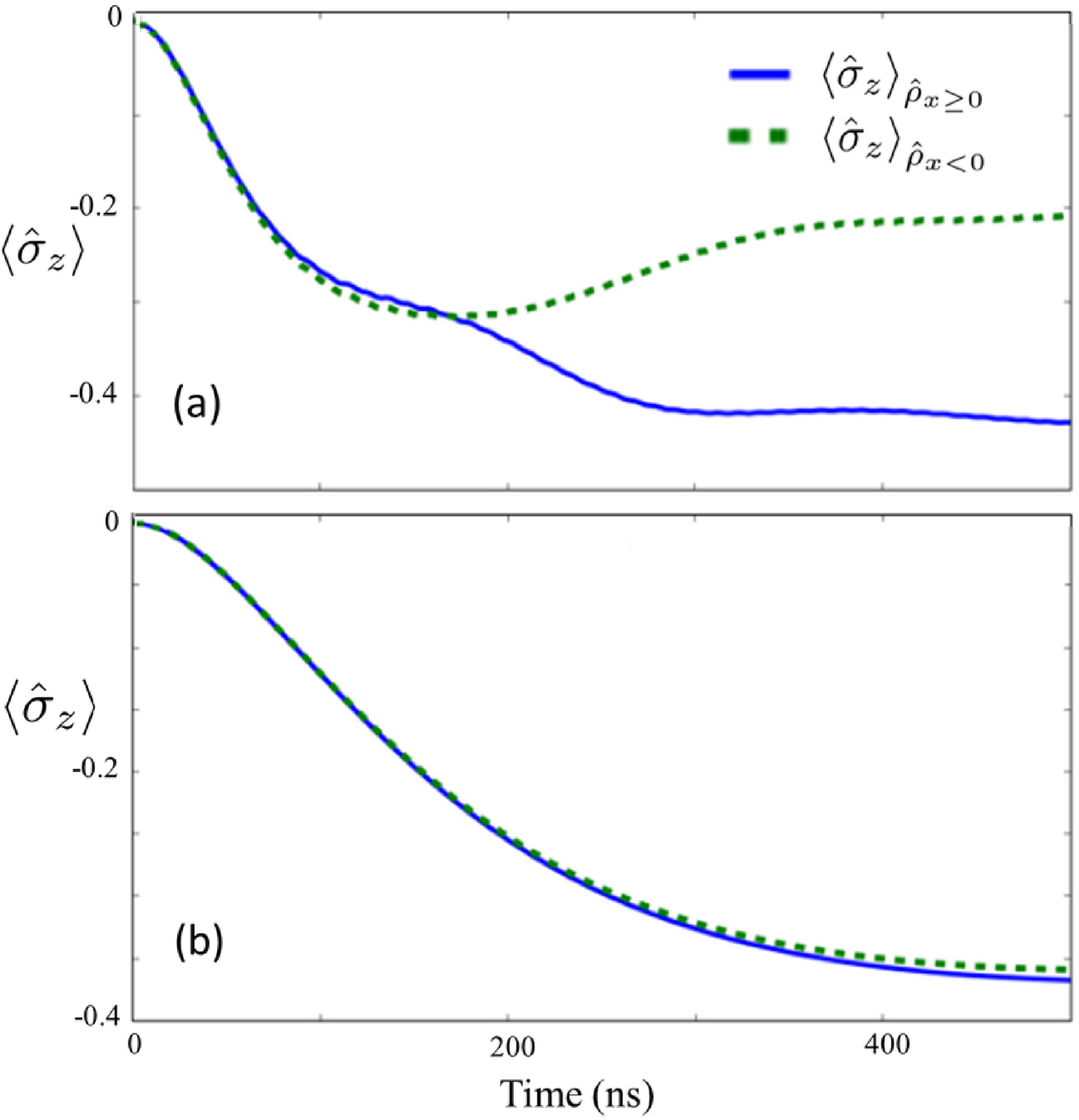}
\caption{(Color online)
 $\braket{\hat{\sigma}_z}$
 after the coarse-graining measurements that projects the
 state into $\hat{\rho}_{x\geq0}$ or $\hat{\rho}_{x<0}$ depending on the measurement
 results. We plot $\braket{\hat{\sigma}_z}$ in (a) for the nonlinear resonator
 and in (b) for the linear resonator as the measurement apparatus.
 Here, we set the coarse-graining value as $\sigma=5$. For the other parameters, we use the
 same as those in
 Fig.\ref{kekka1}.}
\label{kekka3}
\end{figure}

\begin{figure}[t]
\centering
\includegraphics[width=\columnwidth]{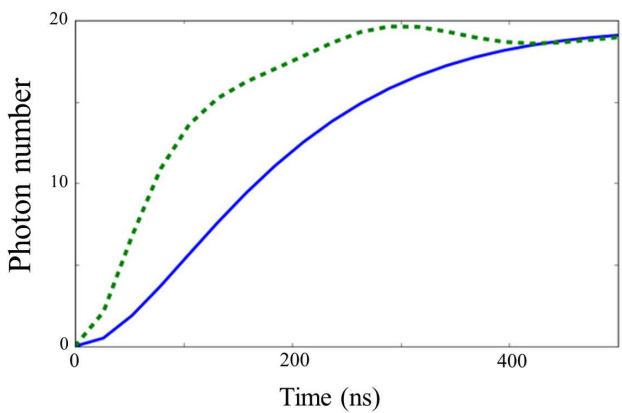}
\caption{(Color online) The average number of photons in the nonlinear resonator (dashed green curve) and in the linear resonator (blue continous curve). The parameters used are the same as those in Fig. \ref{kekka2}.}
\label{photonnumber}
\end{figure}

In Figs.~\ref{kekka2} and \ref{kekka3} we plot the post-measurement observable photons in the USC cavity, $\hat{x}^\prime=(\hat{X}^+ +\hat{X}^-)/2$ and
the state of the qubit $\hat{\sigma}_z=\ket{L}\bra{L}-\ket{R}\bra{R}$, for $\hat{\rho}_{x\geq0}$
and $\hat{\rho}_{x<0}$. For comparison, we consider both a linear resonator ($\chi =0$) and a
nonlinear resonator ($\chi \neq 0$) as the measurement devices. In addition, in Fig. \ref{photonnumber}, we show the average photon number inside the measurement resonator, also for the case of a nonlinear and a linear device.
In all figures, for the nonlinear measurement resonator, when we set the coarse-graining value as $\sigma=5$, the post-measurement state of the USC cavity and qubit changes significantly,
depending on the measurement outcome, and so we
observe a clear measurement backaction on the ultra-strongly coupled
system.
Interestingly, in these examples, we set
the coupling strength $J$ as approximately $300$ times smaller than the
qubit energy.
On the other hand, for a linear resonator, the effect of
the measurement backaction is negligible in this regime, and
the post-measurement state is almost independent of the
measurement results.

\par

\subsection{Low-energy two-level approximation}

\begin{figure}[h!]
\centering
\includegraphics[width=\columnwidth]{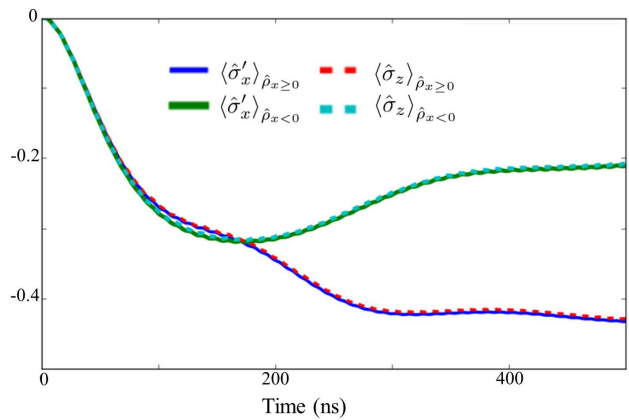}
\caption{\textcolor{black}{(Color online)
 Numerical simulations of
 the expected values of $\hat{\sigma}_x^\prime$ and $\hat{\sigma}_z$ after the
  the state is projected into $\hat{\rho}_{x\geq0}$ or $\hat{\rho}_{x<0}$ depending on the measurement
 results.
 We use the same parameters as those in Fig. \ref{kekka3}.}
 }
\label{twolevelappro}
\end{figure}

To give an intuitive explanation for why
the nonlinear
resonator measurement apparatus can become strongly correlated with the USC system, even when the coupling between measurement apparatus and system is much smaller than
the system energy scales, we introduce a two-level
approximation for the USC system. \textcolor{black}{(see Appendix D for details, and a detailed analysis of the validity of this approximation)}.
In our simulations, the initial state is $\ket{G}$, and the interaction
Hamiltonian $J\hat{\sigma}_z b^\dag
b$ mainly induces a transition from  $\ket{G}$ to
the first excited state
$\ket{E}=\frac{1}{\sqrt{2}}(\ket{R}\ket{\alpha}+\ket{L}\ket{-\alpha})$.
Since the transition matrix elements of the interaction Hamiltonian to
the other excited states are negligible,
we can approximate the low-energy states of the ultra-strongly-coupled system as a two-level
system. In this case,  $\hat{H}_{\rm{Rabi}}$ and $\hat{H}_{\rm{int}}$ can be written as
\begin{align}
\hat{H}_{\rm{Rabi}}& \approx \frac{\omega_{\rm{eff}}}{2} \hat{\sigma}_z^\prime \\
\hat{H}_{\rm{int}}& \approx J \hat{\sigma}_x ^\prime \hat{b}^\dag \hat{b},
\end{align}
where
\begin{equation}
\omega_{\rm{eff}}=\omega_{\rm{q}} \exp[-2\alpha^2],
\end{equation}
and
\begin{align}
\hat{\sigma}_z^\prime&=\ket{E}\bra{E}-\ket{G}\bra{G}\\
\hat{\sigma}_x^\prime&=\ket{G}\bra{E}+\ket{E}\bra{G}.
\end{align}
In Fig.~\ref{twolevelappro}, we plot $\braket{\hat{\sigma}_x^\prime}$
corresponding to $\hat{\rho}_{x\geq0}$ and  $\hat{\rho}_{x<0}$, with this two-level system approximation.
To check the validity of this simplified model, we plot
$\hat{\sigma }_x'$ with this model and $\hat{\sigma}_z$ using the full model in Fig.~\ref{twolevelappro}. These
results show an excellent agreement.

With this two-level approximation, we can
show that the large correlation between the nonlinear
resonator and the ultra-strongly-coupled system originates from the combination of an AC Stark shift and an adiabatic
transition. It is easy to see that the large number of photons in the
nonlinear resonator induces an energy shift (AC Stark shift) of the USC two-level
system. Since the photon number  of
the high-amplitude state is different from that of the low-amplitude
state, the size of the AC Stark shift
strongly depends on the state of the nonlinear resonator.
As long as the timescale of the change in the
nonlinear resonator photons is much smaller than $1/\omega_{\rm{eff}}$,
the state of the two-level system remains in a ground state of the following
effective Hamiltonian 
\begin{align}
\hat{H}_{\rm{eff}}=J\braket{\hat{b}^\dag \hat{b}}_{H(L)} \hat{\sigma}_x^\prime +\frac{\omega_{\rm{eff}}}{2}\hat{\sigma}_z^\prime,
\label{effhamiltonian}
\end{align}
where $\braket{\hat{b}^\dag
\hat{b}}_{H}$($\braket{\hat{b}^\dag \hat{b}}_{L}$) is the average photon
number of the high (low) amplitude state.

\begin{figure}[t]
\centering
\includegraphics[width=\columnwidth]{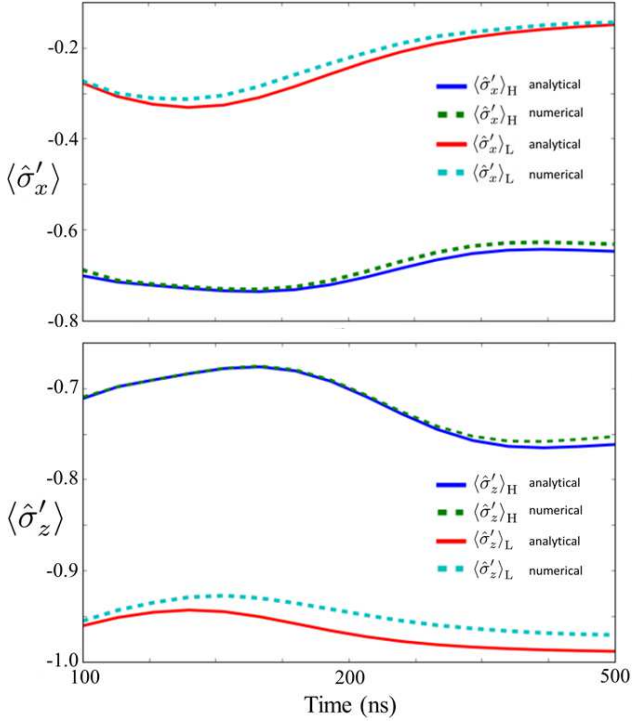}
\caption{(Color online)
Numerical results and analytical solutions of
 the expected values of $\hat{\sigma}_x^\prime$ and $\hat{\sigma}_z^\prime$ after the
  nonlinear resonator is projected into a high-amplitude state or a low-amplitude state.
 In the analytical calculations, we use the simplified Hamiltonian
 described in the Eq.~(\ref{effhamiltonian}).
 }
\label{acstark}
\end{figure}

When the nonlinear measurement resonator becomes a mixed state of
the low- and high-amplitude states, we expect that the AC Stark shift (whose amplitude
depends on the nonlinear resonator state) induces an adiabatic change
of the ground state of the two-level system. This leads to a large
correlation between the USC system and the measurement resonator. To show the validity of this interpretation, we
analytically calculate the $\braket{\hat{\sigma}_{z(x)}^\prime}$ of the ground state
of the Hamiltonian in Eq.~(\ref{effhamiltonian}) where we substitute
the numerically calculated photon numbers of the high (low) amplitude state for $\braket{\hat{b}^\dag
\hat{b}}_{\mathrm{H}}$ ($\braket{\hat{b}^\dag
\hat{b}}_{\mathrm{L}}$).
In Fig.~\ref{acstark}, we compare these results with
the numerical simulations~\cite{qutip,qutip2} where the master equation with the simplified
Hamiltonian is solved. There is a good agreement between these two
results, leading us to conclude that the correlation between the
two-level system and the nonlinear resonator is induced by the
aforementioned adiabatic changes due to the AC Stark shift, whose
amplitude depends
on the nonlinear resonator state. Note that in  Fig. \ref{acstark} we do not show the time
evolution from $t=0\hspace{1mm} \mathrm{ns}$ to $t=100\hspace{1mm}
\mathrm{ns}$, because the high-amplitude state is not generated until
approximately $t=100\hspace{1mm} \mathrm{ns}$.

\par

\subsection{Comparison to QND limit}

\begin{figure}[h!]
\centering
\includegraphics[width=\columnwidth]{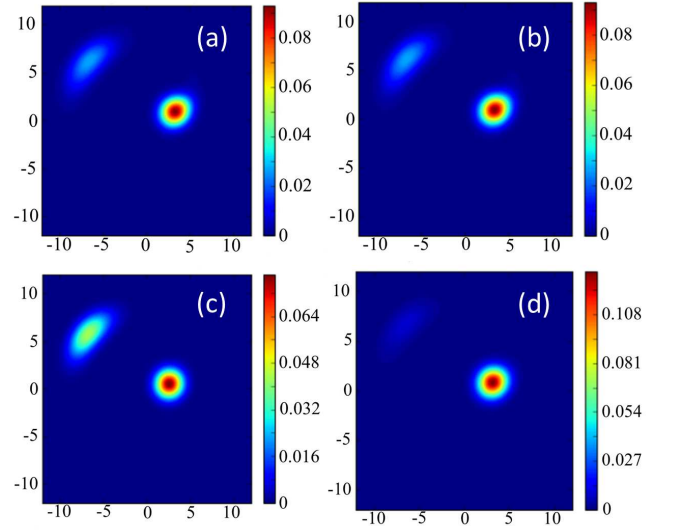}
\caption{(Color online)
 The $Q$ functions of the nonlinear resonator for several conditions:
 (a) Numerical simulation of the
 full Hamiltonian described in Eqs.~(3)-(5). (b) The two-level system approximation. (c)
 Ideal QND measurement, which is possible in the limit $\omega_{\rm{q}}=0$.
 (d) When the nonlinear resonator does not couple at all with the qubit.
We use the same
 parameters as those in Fig. \ref{kekka3}.
 }
\label{qfunctions}
\end{figure}

To compare our non-energy eigenbasis measurements
with a ideal quantum non-demoliton (QND) measurements,
we now study the behavior of the the $Q$ function of the nonlinear resonator, as shown in  Fig.~\ref{qfunctions}.
Here, we consider the following four cases: (a) a non-energy
eigenbasis measurement with the full Hamiltonian described in
Eq.~(\ref{total}), (b) a non-energy
eigenbasis measurement with the two-level system approximation described
by Eq.~(\ref{effhamiltonian}), (c) quantum
non-demoliton measurements for the full Hamiltonian described in
Eq.~(\ref{total}) for the limit $\omega_{\rm{q}}=0$ (which makes the measurement satisfy the QND condition
$[\hat{H}_{\rm{Rabi}},\hat{H}_{\rm{int}}]=0$), and (d) null measurements with $J=0$.

First, we again confirm that the two-level approximation (b) compares well to the full
   Hamiltonian case (a).
 Moreover, we observe a clear difference between our
   non-energy eigenbasis measurements and measurements in the QND limit (c). In particular,
   the probability to obtain the high-amplitude state in the the nonlinear resonator
   becomes much larger for QND measurements than
   that for the the non-energy eigenbasis measurement case.

   Second, a naive application of the rotating-wave approximation to the system and measurement device coupling term, for the non-energy eigenbasis
   measurement case, suggests that the influence of system and measurement apparatus on each other should be entirely negligible. Of course, there is a clear
   difference between the case with a finite $J$ and the case without
   $J$, because such an approximation should also take into account the norm of the operator in the interaction term, which for the driven nonlinear resonator can be large. The figures show that, roughly speaking, the probability to obtain the high-amplitude
   state of the resonator for the non-energy eigenbasis measurements lies
   between the case of the QND measurements and null
   measurements.

\par

\begin{figure}[t]
\centering
\includegraphics[width=\columnwidth]{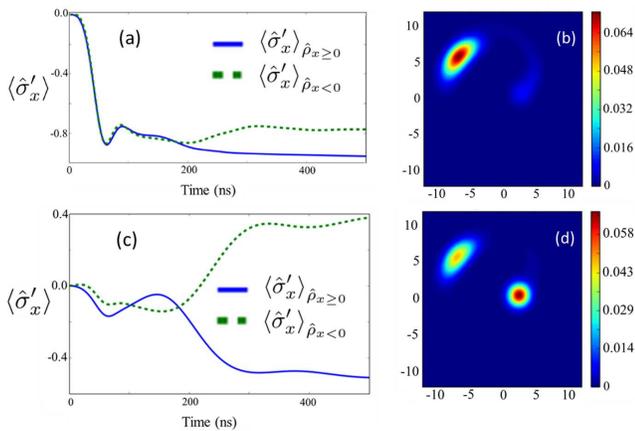}
\caption{(Color online)
 (a,b)~Dynamics of the ultra-strongly-coupled system and the nonlinear
 resonator when the
 effective energy $\omega_{\rm{eff}}$ is 10 times smaller than those in
 Fig.~\ref{kekka3}.
 (a) The expected value of $\hat{\sigma}_x^\prime$ after the
 nonlinear resonator is projected into a high-amplitude state or a low-amplitude state.
 (b) The $Q$ function of the nonlinear resonator at time $500\hspace{1mm} \mathrm{ns} $.
 (c,d) Dynamics of the ultra-strongly-coupled system and the nonlinear
 resonator when the
 effective energy $\omega_{\rm{eff}}$ is 100 times smaller than those in
  Fig.~\ref{kekka3}.
 (c) The expected value of $\hat{\sigma}_x^\prime$ after the
 nonlinear resonator is projected into a high-amplitude state or a low-amplitude state.
 (d) The $Q$ function of the nonlinear resonator at time $500\hspace{1mm} \mathrm{ns} $.
 Except for the effective energy of
the ultra-strongly coupled system, we use the same parameters as
 those in Fig.~\ref{kekka3}. 
}
\label{takusan}
\end{figure}

Also, we increase the ratio $J/\omega_{\rm{q}}$ to check how
the effect of the AC Stark shift will change. In Fig. \ref{takusan}(a), we plot
$\braket{\hat{\sigma}_x^\prime}_{\hat{\rho}_{x\geq 0}} $,
$\braket{\hat{\sigma}_x ^\prime}_{\hat{\rho}_{x<0}}$,
and the $Q$ function at $t=500\hspace{1mm} \mathrm{ns}$
where the effective energy $\omega_{\rm{eff}}$ is $10$\% that used in
Fig.\ref{kekka3}.
From Fig. \ref{takusan}(a), the system converges into an
eigenstate of $\hat{\sigma }_x'$ after the
interaction, regardless of the measurement
results of the nonlinear resonator. This can be understood by considering that the AC
Stark effect $J\langle \hat{b}^{\dagger }\hat{b}\rangle _{\rm{H(L)}}$
becomes much larger than the effective energy $\omega_{\rm{eff}}$  so that the state
of the ultra-strongly-coupled system becomes an eigenstate of
$\hat{\sigma }_x'$ for
both the high amplitude state and low amplitude state.
Furthermore, it is worth mentioning that, from
Fig. \ref{takusan}(b), the nonlinear resonator before the
measurement almost becomes a high-amplitude state. For an ideal quantum
projective measurements on the ground state of the ultra-strongly
coupled system, the population in the low-amplitude state should
be the same as that of the high-amplitude state, and so this result shows
that the effective energy $\omega_{\rm{eff}}$ is still too large to realize
a full projective measurement in the persistent current basis.

We also consider a case when the effective energy $\omega _{\rm{eff}}$ is $1$\% of that used in
Fig.~\ref{kekka3}. In that case, $\braket{\hat{\sigma}_z}_{\hat{\rho}_{x< 0}} $
becomes much larger than
$\braket{\hat{\sigma}_z}_{\hat{\rho}_{x\geq0}}$, and this cannot be
explained just by the AC Stark shift. Moreover, from
Fig.~\ref{takusan}(c), the population of the high-amplitude state
becomes comparable with that of the low-amplitude state. Therefore, in
this regime, we realize a strong projection of the ground state of the ultra-strongly-coupled system in the non-energy eigenbasis.
\section{Negativity}
As a criteria of entanglement, and to understand how correlations between nonlinear resonator and USC system develop, we consider the negativity. Suppose there is a Hilbert space of two systems, $\mathcal{H}_A \otimes \mathcal{H}_B$ with a state $\hat{\rho}_{AB}$. The definition of negativity is
\begin{align}
N(\hat{\rho})=\frac{||\hat{\rho}^{T_A}||-1}{2}
\end{align}
here, $\hat{\rho}^{T_A}$ is the partial transpose of the state $\hat{\rho}_{AB}$ taken over a subsystem $A$, and $||\hat{X}||=\mathrm{Tr}\sqrt{\hat{X}^\dag \hat{X}}$ is the trace norm \cite{negativity1}. In our case, the subsystem $A$ corresponds to the two-level system approximation of the USC system, and $B$ to the nonlinear resonator.
\begin{figure}[t]
\centering
\includegraphics[width=\columnwidth]{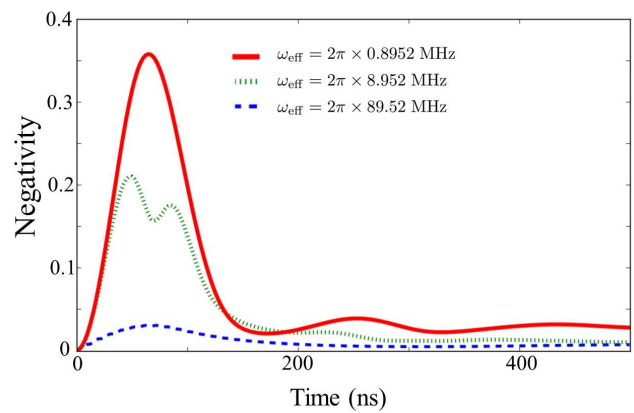}
\caption{(Color online)
Entanglement between the ultra-strongly
 coupled system and the nonlinear resonator. We use the negativity as a
 measure of entanglement.
  From the top, we plot results with effective energies  $\omega_{\rm{eff}}=2\pi \times 0.8952\hspace{1mm} \mathrm{MHz}$, $\omega_{\rm{eff}}=2\pi \times 8.952\hspace{1mm} \mathrm{MHz}$, and
 $\omega_{\rm{eff}}=2\pi \times 89.52\hspace{1mm} \mathrm{MHz}$.
 Except for the effective energy of the ultra-strongly coupled system, here we use the same parameters as
 those in Fig. \ref{kekka3}.
 }
\label{negativity}
\end{figure}
\textcolor{black}{In Fig.~\ref{negativity}} we plot the negativity to quantify the
entanglement between the ultra-strongly-coupled system and the
nonlinear resonator. As we increase the ratio $J/\omega_{\rm{q}}$, the
negativity also increases. These results show that a reasonably large entanglement
between the ultra-strongly-coupled system and the nonlinear resonator
is generated in the regime where we realize a projective measurement
on the non-energy eigenbasis. However, due to the decoherence of the
nonlinear resonator, the entanglement quickly degrades, and a classical
correlation remains in these systems just before the measurement on the nonlinear resonator.
\par
\begin{figure}[t]
\centering
\includegraphics[width=\columnwidth]{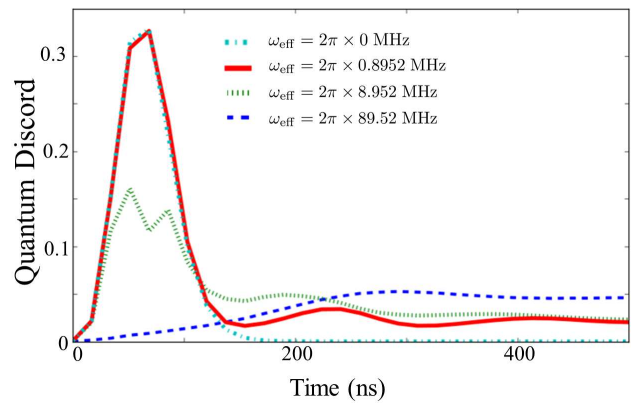}
\caption{(Color online)
Quantum discord between the ultra-strongly
 coupled system and the nonlinear resonator.
  From the top, we plot results with effective energies $\omega_{\rm{eff}}=2\pi \times 0\hspace{1mm} \mathrm{MHz}$, $\omega_{\rm{eff}}=2\pi \times 0.8952\hspace{1mm} \mathrm{MHz}$, $\omega_{\rm{eff}}=2\pi \times 8.952\hspace{1mm} \mathrm{MHz}$, and
 $\omega_{\rm{eff}}=2\pi \times 89.52\hspace{1mm} \mathrm{MHz}$.
 Except for the effective energy of the ultra-strongly coupled system, here we use the same parameters as
 those in Fig. \ref{kekka3}.
 }
\label{discord}
\end{figure}
\section{Quantum discord}
To elucidate  the previous results further, we consider the quantum discord (QD), which is defined as follows.  Two possible definitions of the mutual information of the state $\hat{\rho}_{AB}$
\begin{align}
I(\hat{\rho}_{AB})&=S(\hat{\rho}_A)+S(\hat{\rho}_B)-S(\hat{\rho}_{AB}) \\
J_A(\hat{\rho}_{AB})&=S(\hat{\rho}_B)-S(\hat{\rho}_B|\hat{\rho}_A)
\end{align}
where $S(\hat{\rho})$ is a von Neumann entropy for a state $\hat{\rho}$, $\hat{\rho}_{A(B)}$ is a reduced density operator for $\mathcal{H}_{A(B)}$,  and $S(\hat{\rho}_B|\hat{\rho}_A)$ is a quantum generalization of a conditional entropy. In the purely classical case, one can show that these two definitions of the mutual information are equivalent. However, in the nonclassical case, these definitions do not necessarily coincide. Also, $J_A(\hat{\rho}_{AB})$ is dependent on the measurement basis $\hat{M}^A$ for  $\mathcal{H}_A$. Therefore, QD is defined as
\begin{align}
\mathcal{Q}&=I(\hat{\rho}_{AB})-\mathrm{max}_{\hat{M}^A} \{J_{\hat{M}^A} (\hat{\rho}_{AB})\} \\
&=S(\hat{\rho}_A)-S(\hat{\rho}_{AB})+\mathrm{min}_{\hat{M}^A} S(\hat{\rho}_{B|\{\hat{M}^A\}})
\end{align}
where 
\begin{equation}
S(\hat{\rho}_{B|\{\hat{M}^A\}})=\sum_k p_k S(\hat{M}_k^A\hat{\rho}_{AB} \hat{M}_k^A/p_k),
 \end{equation}
and 
\begin{equation}
p_k=\mathrm{Tr}(\hat{M}_k^A \hat{\rho}).
 \end{equation} 
 Here $\hat{M}_k^A$ is a  projector when the result is $k$, and QD is basis independent and reflects only nonclassical correlations~\cite{discord1,discord2}.   In our case, system $A$ corresponds to the approximated two level system and system $B$ the nonlinear resonator. We set the measurement basis on the approximated two level system as $\{\ket{\varphi_1}\bra{\varphi_2}, \ket{\varphi_2}\bra{\varphi_2}\}$, 
 \begin{align}
 \ket{\varphi_1}&=\mathrm{cos}(\theta/2)\ket{g}+e^{i\phi}\mathrm{sin}(\theta/2)\ket{e},\\ 
 \ket{\varphi_1}&=\mathrm{sin}(\theta/2)\ket{g}-e^{i\phi}\mathrm{cos}(\theta/2)\ket{e},
 \end{align}
  $(0\leq \theta\leq \pi, 0 \leq \phi <2\pi )$, where $\ket{e}$ and $\ket{g}$ are the eigenstates of $\hat{\sigma}_x^\prime$. Given these definitions we find the $(\theta, \phi)$ which realizes $\mathrm{min}_{\hat{M}^A} S(\hat{\rho}_{B|\{\hat{M}^A\}})$.
\par
\textcolor{black}{We plot the QD in Fig.~\ref{discord}.
Interestingly, in contrast to the negativity, the QD, at $t=500
\hspace{1mm} \mathrm{ns} $, becomes larger as $J/\omega_{\rm{q}}$
is decreased.
This can be explained in the following way: if $J/\omega_{\rm{q}}$ is sufficiently
large, the state becomes a highly entangled state well approximated by the form
\begin{equation}
\frac{1}{\sqrt{2}}(\ket{e}\ket{\rm{Low}}-\ket{g}\ket{\rm{High}}),
\end{equation}
which decays, due the measurement of the nonlinear cavity, to the mixture
\begin{equation}
\hat{\rho}_{\rm{f}}=\frac{1}{2}(\ket{e}\bra{e}\otimes\ket{\rm{Low}}\bra{\rm{Low}}+\ket{g}\bra{g}\otimes\ket{\rm{High}}\bra{\rm{High}}),
\end{equation}
where $\ket{\rm{High}}$ and $\ket{\rm{Low}}$ are high and low amplitude
states of the nonlinear resonator.
 \textcolor{black}{Since $\ket{e}$ and $\ket{g}$ are} orthogonal to each
 other, $\hat{\rho}_{\rm{f}}$ is a classically correlated state
 \textcolor{black}{without any superposition}, implying vanishing QD. }
\textcolor{black}{On the other hand, when $J/\omega_{\rm{q}}$ is small,
the dynamics can be explained by an AC Stark shift and the state can be
expressed as
\begin{eqnarray}
\hat{\rho}_{\mathrm{a}}&=&p_{\rm{H}}
\ket{\psi_{\rm{H}}}\bra{\psi_{\rm{H}}}
\otimes\ket{\rm{High}}\bra{\rm{High}}\nonumber\\&+&p_{\rm{L}}
\ket{\psi_{\rm{L}}}\bra{\psi_{\rm{L}}}
\otimes\ket{\rm{Low}}\bra{\rm{Low}},
\end{eqnarray}
where $p_{\rm{L}(\rm{H})}$ is the
probability that the nonlinear resonator is in the low (or high) amplitude
state. The state $\ket{\psi_{\rm{H}(\rm{L})}}$ is the ground state of
$\hat{H}_{\rm{eff}}$. Here, $\ket{\psi_{\rm{H}}}$  and
$\ket{\psi_{\rm{L}}}$ are not
\textcolor{black}{always}
orthogonal to each other, and as such the correlation in the mixture of the two \textcolor{black}{could
have} a non-classical nature. Hence, the QD, in the long-time limit, tends to have a finite value when \textcolor{black}{
$J/\omega_{\rm{q}}$ is small.}}
\par
\vspace{5mm}

\section{Measurement of initial states not in the energy eigenbasis}
\textcolor{black}{
Conventionally, in evaluating the performance of a readout device, one considers how well the final state of the measurement device correlates with the different possible initial states of the system, as discussed in \cite{trajectory,clerk1}.  In our case, this conventional approach does not reveal sufficient information about how well one can project something like the ground state of a USC system onto a non-eigenstate. }

\begin{figure}[t]
\centering
\includegraphics[width=\columnwidth]{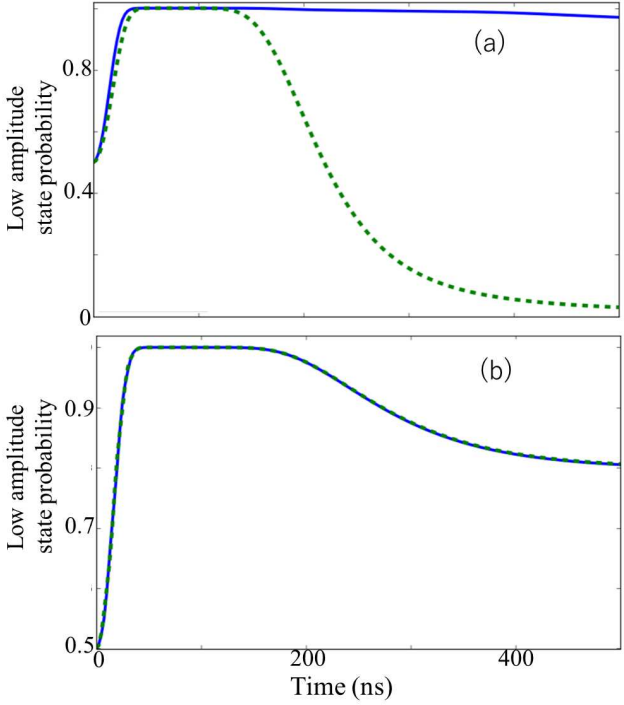}
\caption{Plot of the low-amplitude state probability for different initial states of the USC system, corresponding to eigenstates of the $\hat{\sigma}^\prime_x$  in the two-level approximation. The blue curve corresponds to the case where we set the initial state to $\ket{g} = \ket{L} \ket{-\alpha}$ and the dashed green curve to $\ket{e} = \ket{R}\ket{\alpha}$. In (a) we set $\omega_{\rm{eff}} = 2\pi \times 0.8952~\rm{MHz}$ and hence $\omega_{\rm{eff}}/J = 0.9425$. In (b) we set $\omega_{\rm{eff}} = 2\pi \times 89.52~\rm{MHz}$ and hence $\omega_{\rm{eff}} /J = 94.25$. All other parameters are the same as in Fig 2.}
\label{cor}
\end{figure}

\textcolor{black}{For example, in Fig.~\ref{cor}  we plot the probability
for the resonator to be a low amplitude state depending on the initial
states of the USC system.  In particular, we choose these different states to be not eigenstates of the system Hamiltonian but eigenstates of the $\sigma'_x$ operator (in the two-level system approximation) which couples to the measurement device.  These eigenstates correspond to $\ket{e}=\ket{R} \ket{\alpha}$ and $\ket{g}=\ket{L} \ket{\alpha}$  in the full basis, as described in the previous section. From these graphs, we can see that in
the regime $\omega_{\rm{eff}}/J = 94.25$ [Fig.~\ref{cor}(b)], there is no correlation between the state of
the nonlinear resonator and the initial state of system. On the other hand, for much stronger couplings between system and measurement device,
$\omega_{\rm{eff}}/J = 0.9425 $ [Fig.~\ref{cor}(a)], there is a strong correlation between the nonlinear resonator
and the initial state of the system. }

\textcolor{black}{
These figures suggest that, in this conventional picture, the coupling strength should
be comparable with the effective energy of the system to realize a measurement of an initial state which is not in an eigenstate of the Hamiltonian. This is because, when the initial state is not such an eigenstate, the qubit evolves under the system Hamiltonian with a
time scale corresponding to the inverse of the eigenenergy (in this case, $\omega_{\rm{eff}}$). If the interaction between the
system and measurement apparatus is much weaker than the eigenenergy of
the system Hamiltonian, the system initial state evolves
before the measurement apparatus obtains information about that initial
state, and so the non-linear resonator has no time to build a correlation with the
initial state of the system. }

\textcolor{black}{
On the other hand, if the system is prepared in the
ground state of the system Hamiltonian, the system does not, initially, evolve under the system Hamiltonian. This gives time for the the measurement apparatus to build up significant amount of photons, and become correlated with the system, due to the AC Stark shift, even in the regime of $\omega_{\rm{eff}}/J = 94.25$, as shown in the earlier sections of this work.}

\section{Conclusions}
In conclusion, we investigated quantum measurements in the ultra-strong-coupling
regime of a light-matter system.
In particular, we showed how the ground state of an ultra-strongly-coupled system can be measured by a nonlinear resonator.
Interestingly, we found that, even if the coupling strength with the measurement device is two orders of
magnitude smaller than the typical energy scale of the
ultra-strongly-coupled system, we can still induce a strong classical
correlation with the measurement device.
Also, we confirmed that, by increasing the coupling strength with the
measurement device, entanglement between the system and measurement
device can be generated, and we can
realize projective measurements on the ground state of the
ultra-strongly-coupled system. \textcolor{black}{ In addition, we found that the quantum discord tends to have a finite value at large times in the regime when the dynamics can be described by AC Stark shift.}
Our results help illuminate the mechanism of how an
ultra-strongly coupled system interacts with a measurement device.

\vspace{3mm}
\acknowledgements
We acknowledge helpful discussions with R. Stassi and H. Toida.
This work was supported by JSPS KAKENHI Grants
15K17732 and MEXT KAKENHI
 Grant Number 15H05870. FN acknowledges support from the MURI Center for
Dynamic Magneto-Optics via the AFOSR Award No.
FA9550-14-1-0040, the Japan Society for the Promotion
of Science (KAKENHI), the IMPACT program of JST,
JSPS-RFBR grant No 17-52-50023, CREST grant No.
JPMJCR1676.
NL and FN acknowledge support by the Sir John Templeton
Foundation and the RIKEN-AIST Joint Research Fund.

\section*{Appendix A: Non-energy-eigenbasis measurements}
 \textcolor{black}{Here, we explain the reason why the non-energy
 eigenbasis measurement is difficult to realize.}
 Naive calculations indicate that the non-energy
eigenbasis measurements would require a violation of the rotating wave
approximation, which needs a strong coupling between the system and apparatus. This seems to suggest that, unless the coupling between the
system and measurement apparatus is as large as the resonant frequency of
the system and measurement apparatus, it would be difficult to implement
the non-energy basis measurements. However, our results show that this
naive picture is actually wrong if we use the non-linear resonator as a
measurement apparatus.

We can explain these points more quantitatively as follows. Suppose the Hamiltonian which expresses the coupling between a qubit and a linear resonator as follows.
\begin{align}
\textcolor{black}{\hat{H}=\frac{\omega_{\mathrm{eff}}}{2}\hat{\sigma}_z^\prime+J \hat{\sigma}_x^\prime\hat{b}^\dag \hat{b}+\omega_{\mathrm{r}}\hat{b}^\dag \hat{b} }
\end{align}
\par
\textcolor{black}{In a rotating frame defined by a unitary operator $\hat{U}=\mathrm{exp}\big[-i (\omega_{\mathrm{eff}}\hat{\sigma}_z/2+\omega_{\mathrm{r}}\hat{b}^\dag \hat{b} )t  \big]$, we obtain }
\begin{align}
\hat{H}(t)=J(\mathrm{exp}[i \omega_{\mathrm{eff}} t] \hat{\sigma}_+^\prime +\mathrm{exp}[-i \omega_{\mathrm{eff}} t ] \hat{\sigma}_-^\prime )\hat{b}^\dag \hat{b}
\end{align}
\par
\textcolor{black}{In the limit of a large $\omega_{\mathrm{eff}}$, we can use a rotating wave approximation and we obtain}
\begin{align}
\textcolor{black}{\hat{H}(t) \approx 0}
\end{align}
\textcolor{black}{in the rotating frame.}
\par
\textcolor{black}{More generally, we have a hamiltonian}
\begin{align}
\textcolor{black}{\hat{H}=\hat{H}_{\mathrm{S}}+J \hat{A} \otimes \hat{B} +\hat{H}_{\mathrm{E}} }
\end{align}
\textcolor{black}{where $\hat{H}_{\mathrm{S}}=\sum_n E_n^{(\mathrm{S})} \ket{E_n^{(\mathrm{S})}}\bra{E_n^{(\mathrm{S})}}$ and  $\hat{H}_{\mathrm{E}}=\sum_m E_m^{(\mathrm{E})} \ket{E_m^{(\mathrm{E})}}\bra{E_m^{(\mathrm{E})}}$, (the superindex (S) denotes the system and the superindex (E) denotes the measurement apparatus). In a rotating frame defined by $\hat{U}=\mathrm{exp}[-it(\hat{H}_{\mathrm{S}}+\hat{H}_{\mathrm{E}})]$, we have }
\begin{align}
\hat{H}(t)&= J \sum_{n,n^\prime,m,m^\prime} C_{n,n^\prime,m,m^\prime} \ket{E_n^{(\mathrm{S})}}\bra{E_{n^\prime}^{(\mathrm{S})}} \otimes \ket{E_m^{(\mathrm{E})}}\bra{E_{m^\prime}^{(\mathrm{E})}} \\ \nonumber
& \times \mathrm{exp}[-i (E_n^{(\mathrm{S})}-E_{n^\prime}^{(\mathrm{S})})t -i (E_m^{(\mathrm{E})}-E_{m^\prime}^{(\mathrm{E})})t]
\end{align}
\textcolor{black}{where $C_{n,n^\prime,m,m^\prime}=\bra{E_n^{(\mathrm{S})}}\hat{A}\ket{E_{n^\prime}^{(\mathrm{S})}} \bra{E_m^{(\mathrm{E})}}\hat{B}\ket{E_{m^\prime}^{(\mathrm{E})}}  $. If the system and measurement apparatus are well detuned, we obtain }
\begin{align}
\textcolor{black}{\hat{H}(t) \approx \sum_{n,m} C_{n,n,m,m} \ket{E_n^{(\mathrm{S})}}\bra{E_{n}^{(\mathrm{S})}} \otimes \ket{E_m^{(\mathrm{E})}}\bra{E_{m}^{(\mathrm{E})}}}
\end{align}
\textcolor{black}{where we used the rotating wave approximation. So the terms that commute with $\hat{H}_{\mathrm{S}}$ survive. This clearly shows that we can measure only an observable that commutes with $\hat{H}_{\mathrm{S}}$ if the rotating wave approximation is valid. This also means that we need a violation of the rotating wave approximation for the non-energy eigenbasis measurements. }

\section*{Appendix B:  Derivation of the interaction Hamiltonian between the nonlinear resonator and the qubit}
\textcolor{black}{ In this work we rely on an interaction between a superconducting flux
qubit coupled with a frequency tunable resonator.  This is not a dispersive approximation to a dipolar coupling.
In more detail, the flux qubit is described as}
\begin{eqnarray}
 \textcolor{black}{\hat{H}_{\rm{fq}}=\frac{\epsilon }{2}\hat{\sigma }_z +\frac{\Delta }{2}\hat{\sigma }_x}
\end{eqnarray}
\textcolor{black}{where $\epsilon $ denotes an energy bias and $\delta $ denotes a tunneling
energy. The Pauli matrix $\hat{\sigma }_z$ denotes a population of a
persistent current basis such as $\hat{\sigma }_z=|L\rangle \langle L|-
|R\rangle \langle R|$ where $|L\rangle $ $|R\rangle$ denotes a left-sided (right-handed) persistent
current.}
\par
\textcolor{black}{The frequency tunable resonator is described as}
\begin{eqnarray}
 \textcolor{black}{\hat{H}_r=\omega (\Phi)\hat{a}^{\dagger }\hat{a}}
\end{eqnarray}
\textcolor{black}{where $\omega (\Phi)$ denotes a frequency of the resonator. We assume
that the resonator contains a SQUID structure, and we can tune the
frequency of the resonator by changing an applied flux penetrating the
SQUID structure.
(For example, see \cite{reso1}).}
\par
\textcolor{black}{We can derive the interaction between the flux qubit and resonator as
follows. The persistent current states of the flux qubit induces
magnetic fields due to the Biot-Savart law, and this changes the penetrating
magnetic flux of the SQUID in the resonator. So the frequency of the
resonator depends on the state of the flux qubit.
Suppose that $\delta \Phi $ ($-\delta \Phi $) denotes the magnetic flux
from the $|L\rangle $ ($|R\rangle $) state, and the resonator frequency will be approximately shifted by
$\frac{d\omega }{d\Phi }\delta \Phi $ ($-\frac{d\omega }{d\Phi }\delta \Phi $).
This provides us with the following Hamiltonian.}
\begin{align}
\textcolor{black}{\hat{H}_I}&=\textcolor{black}{g |L\rangle \langle L|\otimes \hat{a}^{\dagger }\hat{a}-g
  |R\rangle \langle R|\otimes \hat{a}^{\dagger }\hat{a}} \nonumber \\
&\textcolor{black}{=g \hat{\sigma }_z\otimes \hat{a}^{\dagger }\hat{a}}
\end{align}
\textcolor{black}{where $g=\frac{d\omega }{d\Phi }\delta \Phi$.
A similar Hamiltonian has been derived in \cite{reso2}
to represented a coupling between an NV center and flux qubit.}
\par
\textcolor{black}{Note that we assume a large detuning between the flux
qubit and resonator. In this case dipolar coupling  is negligible.}
\par

\section*{Appendix C:  Derivation of the coarse graining measurement}
\textcolor{black}{In the case that there is noise in the measurement apparatus, when we have a position measurement, even if the result of the measurement apparatus is $x$, the real value is not necessarily $x$. To model such situations, we define a measurement operator as follows}
\begin{align}
\textcolor{black}{\hat{E}_x=\frac{1}{\pi^{1/4}\sqrt{2\sigma}}\int _{-\infty}^{\infty}dx^\prime \mathrm{exp}\bigg[-\frac{(x^\prime-x)^2}{4\sigma^2}\bigg]\ket{x^\prime}\bra{x^\prime}}
\end{align}
\textcolor{black}{where $\sigma$ implies the strength of the noise. $\hat{E}_x$ satisfies the normalization condition}
\begin{align}
\textcolor{black}{\int_{-\infty}^{\infty}\hat{E}_x^\dag \hat{E}_x dx =I }
\end{align}
\textcolor{black}{Here, we consider a composite system which comprises of a system which we hope to readout~(ultra-strongly coupled system) and its probe~(nonlinear resonator). Also, the measurement result is divided to $x\geq0$ and $x<0$. When we have a measurement on a composite system $\hat{\rho}$, the post measurement state  when the result is $x\geq0$   becomes}
\begin{align}
&\textcolor{black}{\frac{\int_0^\infty dx \hat{E}_x\hat{\rho}\hat{E}_x^\dag}{\mathrm{Tr}[\int_0^\infty dx \hat{E}_x\hat{\rho}\hat{E}_x^\dag]} }
\textcolor{black}{=\frac{1}{N}\int _0^\infty dx \int_{-\infty}^{\infty} dx^\prime \int _{-\infty}^{\infty} dx^{\prime \prime} }\\ \nonumber
 &\mathrm{exp}\bigg[
-\frac{(x^\prime-x)^2}{4\sigma^2}-\frac{(x^{\prime \prime}-x)^2}{4\sigma^2} \bigg]
\textcolor{black}{\times \bra{x^\prime} \hat{\rho} \ket{x^{\prime \prime}} \ket{x^\prime} \bra{x^{\prime \prime}}}
\end{align}
\textcolor{black}{where 
\begin{equation}
N=\int_0 ^\infty dx \int_{-\infty}^\infty dx^\prime \mathrm{exp}[-\frac{(x^\prime-x)^2}{2\sigma^2}]\mathrm{Tr}[\bra{x^\prime} \hat{\rho} \ket{x^\prime}].
\end{equation}
By tracing out the probe system, we have the post measurement state of the system $\hat{\rho}_{x\geq 0}$ we hope to readout as}
\begin{align}
\textcolor{black}{\hat{\rho}_{x\geq 0}=\frac{1}{N } \int_0^\infty dx~ \int_{-\infty}^\infty dx^\prime~ \mathrm{exp}\bigg[\frac{(x^\prime-x)^2}{2\sigma^2} \bigg] \bra{x^\prime}\hat{\rho}\ket{x^\prime} }
\end{align}
\textcolor{black}{Substituting $t= \frac{x-x^\prime}{\sqrt{2}\sigma}$, $\hat{\rho}_{x\geq 0}$ can be rewritten as }
\begin{align}
\textcolor{black}{\hat{\rho}_{x\geq 0}=\frac{1}{N} \int_{-\infty} ^{\infty} dx \hspace{1mm} \mathrm{erfc} \bigg( -\frac{x}{\sqrt{2}\sigma} \bigg) \bra{x}\hat{\rho} \ket{x}}
\end{align}
\textcolor{black}{where $\mathrm{erfc}(x)$ is a complementary error function, and is defined as}
\begin{align}
\textcolor{black}{\mathrm{erfc}(x)=\frac{2}{\sqrt{\pi}} \int _x ^\infty \mathrm{exp}(-t^2) dt}
\end{align}
\textcolor{black}{In the limit of $\sigma \rightarrow +0$, we have}
\begin{align}
\textcolor{black}{\hat{\rho}_{x\geq 0}=\frac{1}{N} \int_0^\infty dx \bra{x}\hat{\rho}\ket{x} }
\end{align}
\textcolor{black}{which is noiseless measurement. Also, in the limit of $\sigma \rightarrow +\infty$, we obtain}
\begin{align}
\textcolor{black}{\hat{\rho}_{x\geq 0}=\frac{1}{N} \int_{-\infty}^\infty dx \bra{x}\hat{\rho}\ket{x} }
\end{align}
\textcolor{black}{which shows we cannot have any information from the system.}

\section*{Appendix D:  Validity of the two level approximation}
\subsection{Adiabatic approximation to the Rabi Hamiltonian}
 \textcolor{black}{We now explain the adiabatic approximation to the
 Rabi Hamiltonian, that has also been used in previous works
 \cite{ultra1,ultra2,ultra3,ultra4}. We will show
 that, within the framework of the adiabatic approximation, the
 ultra-strongly coupled system can be treated as a two-level system.}
\textcolor{black}{The conventional Rabi Hamiltonian can be written as}
\begin{align}
\textcolor{black}{\hat{H}_{\mathrm{Rabi}}=\frac{\omega_{\mathrm{q}}}{2}\hat{\sigma}_x+g(\hat{a}+\hat{a}^\dag)\hat{\sigma}_z+\omega_{\mathrm{r}}\hat{a}^\dag\hat{a}}
\label{Rabi}
\end{align}
\textcolor{black}{The adiabatic approximation can be done when $\omega_q \ll (g, \omega_r)$ and the Rabi Hamiltonian can be diagonalized using the bases}
\begin{align}
\textcolor{black}{\ket{L}\ket{N_-}}&\textcolor{black}{=\ket{L}\hat{D}(-\alpha) \ket{N}} \\
\textcolor{black}{\ket{R}\ket{N_+}}&\textcolor{black}{=\ket{R}\hat{D}(\alpha) \ket{N}} \\
\textcolor{black}{\alpha}&\textcolor{black}{=g/\omega_{\mathrm{r}}}
\end{align}
\textcolor{black}{where $\ket{L}$ and $\ket{R}$ are eigenstates of $\hat{\sigma}_z$, $\ket{N}$ is the level of the eigenstates of $\hat{a}^\dag \hat{a}$, and $\hat{D}(\alpha)$ is a displacement operator. The states  $\ket{L}\hat{D}(-\alpha) \ket{N}$ and $\ket{R}\hat{D}(\alpha) \ket{N} $ are degenerate in energy and their energy is $\mathcal{E}_N=\omega_{\mathrm{r}}(N-\alpha^2)$. Then, considering that the term $\omega_q/2 \hat{\sigma}_x$  couples these terms, and only the transitions between the states of the same $N$ are taken into account in the adiabatic approximation, the Rabi hamiltonian can be rewritten as}
\begin{align}
\textcolor{black}{\hat{H}_{\mathrm{Rabi}}\approx \sum_{N=0} \bigg[(\mathcal{E}_N+\frac{\omega_{\mathrm{q}}}{2}\braket{N_-|N_+})\ket{\psi^+_N}\bra{\psi^+_N}}  \\ \nonumber
+(\mathcal{E}_N-\frac{\omega_{\mathrm{q}}}{2}\braket{N_-|N_+})\ket{\psi ^-_N}\bra{\psi^-_N}  \bigg]
\end{align}
\textcolor{black}{where}
\begin{align}
\textcolor{black}{\ket{\psi_N^\pm}=\frac{1}{\sqrt{2}}(\ket{L}\ket{N_-}\pm\ket{R}\ket{N_+}) }
\end{align}
whose eigenvalues are
\begin{align}
\mathcal{E}_{N\pm}=\mathcal{E}_N\pm\frac{\omega_{\mathrm{q}}}{2}\braket{N_-|N_+}
\end{align}
\par
\textcolor{black}{Also, it can be easily shown}
\begin{align}
\textcolor{black}{\bra{\psi_N^\pm}\hat{\sigma}_z\ket{\psi_M^\pm}}&\textcolor{black}{=0} \\
\textcolor{black}{\bra{\psi_N^\mp}\hat{\sigma}_z\ket{\psi_M^\pm}}&\textcolor{black}{=\delta_{NM}}
\end{align}
\textcolor{black}{So, as long as we apply the adiabatic approximation, the transition due to the $\hat{\sigma}_z$ term is between $\ket{\psi_N^+}$ and $\ket{\psi_N^-}$. Since the interaction between the ultra-strongly coupled system and the non-linear resonator can be expressed as $J \hat{\sigma}_z \hat{a}^\dag \hat{a}$, it is possible for us to consider that the ultra-strongly coupled system is driven only by the $\hat{\sigma}_z$ operator. Also, if the initial state is $\ket{\psi_0^-}$ and the perturbation term is proportional only to $\hat{\sigma}_z$ (which applies to our system, which is composed of a ultra-strongly coupled system and a nonlinear resonator, where the interaction term can be expressed as $J\hat{\sigma}_z \hat{a}^\dag \hat{a}$), the dynamics is limited to $\ket{\psi^\pm_0}$. }\textcolor{black}{Therefore, as long as the adiabatic approximation is
valid, we can consider our system of the ultra-strongly coupled
system as a two-level system.}
\subsection{Estimation of the deviation from the two-level approximation.}
\textcolor{black}{By calculating the deviation from the
two-level approximation, we show a quantitative analysis how accurate
the two-level system approximation is in our parameter regime. }\textcolor{black}{ We consider a fidelity between the true ground state $\ket{G}$ (the first excited state $\ket{E}$ ) and $ \ket{\psi_0^-}$ ($\ket{\psi_0^+}$.)}  \textcolor{black}{It is possible to estimate the
accuracy of our two-level approximation from this fidelity, and we derive a
condition of the fidelity to be close to the unity.}
\textcolor{black}{Now, we define}
\begin{align}
\textcolor{black}{\hat{H}_0= \sum_{N=0} \bigg[(\mathcal{E}_N+\frac{\omega_{\mathrm{q}}}{2}\braket{N_-|N_+})\ket{\psi^+_N}\bra{\psi^+_N} } \\ \nonumber
+(\mathcal{E}_N-\frac{\omega_{\mathrm{q}}}{2}\braket{N_-|N_+})\ket{\psi ^-_N}\bra{\psi^-_N}  \bigg]
\end{align}
\textcolor{black}{and}
\begin{align}
\textcolor{black}{\hat{H}^\prime=\hat{H}_{\mathrm{Rabi}}-\hat{H}_0 }
\end{align}
\textcolor{black}{Here, $\hat{H}_{\mathrm{Rabi}}$ is the one defined in Eq.~\ref{Rabi}. In this way, we regard $\hat{H}_0$ as the non-perturbative Hamiltonian and $\hat{H}^\prime$ the perturbative Hamiltonian. }\textcolor{black}{By performing a perturbative calculation up to the lowest order, we obtain}
\begin{align}
\textcolor{black}{\ket{G}\approx \frac{1}{\mathcal{\sqrt{N}}}(\ket{\psi_0^-}+\ket{G^{(0)}})}
\end{align}
\textcolor{black}{where $\mathcal{N}$ is a normalization factor. Then by using perturbation theory, we have}
\begin{align}
\textcolor{black}{\ket{G^{(0)}}}&\textcolor{black}{=\sum_N (c_N^+ \ket{\psi_N^+}+c_N^- \ket{\psi_N^-})}\\
\textcolor{black}{c_N^+}&\textcolor{black}{=-\frac{\bra{\psi_N^+}\hat{H}^\prime\ket{\psi_0^-} }{\mathcal{E}_{N+}-\mathcal{E}_{0-}}} \\
\textcolor{black}{c_N^-}&\textcolor{black}{=-\frac{\bra{\psi_N^-}\hat{H}^\prime\ket{\psi_0^-} }{\mathcal{E}_{N-}-\mathcal{E}_{0-}}}
\end{align}
 \textcolor{black}{In the perturbative calculation, the eigenstate after adding the
perturbative term is not normalized to unity, and so we consider a normalization
factor $\mathcal{N}$ for $\ket{G}$ such as
\begin{equation}\ket{G}\approx
 \frac{1}{\mathcal{\sqrt{N}}}(\ket{\psi_0^-}+\ket{G^{(0)}}).
 \end{equation}} \textcolor{black}{It can be easily shown that 
 \begin{equation} \bra{\psi_N^+}\hat{H}^\prime\ket{\psi_0^-} =\bra{\psi_N^+}\hat{H}_{\mathrm{Rabi}}\ket{\psi_0^-}=\omega_{\mathrm{q}}/2\bra{\psi_N^+}\hat{\sigma}_x\ket{\psi_0^-}
 \end{equation}
and \begin{equation} \bra{\psi_N^-}\hat{H}^\prime\ket{\psi_0^-}=\omega_{\mathrm{q}}/2\bra{\psi_N^-}\hat{\sigma}_x\ket{\psi_0^-}.\end{equation} Then,  we have }
\begin{align}
\textcolor{black}{\ket{G}}& \approx\frac{1}{\sqrt{\mathcal{N}}} \bigg(\ket{\psi_0^-} \\ \nonumber
&+\frac{\omega_{\mathrm{q}}}{2}\sum_{N=2,4,..}\frac{(2\alpha)^N}{(\mathcal{E}_{N-}+\omega\alpha^2+\omega_{\mathrm{q}}/2\braket{0-|0+})\sqrt{N!}} \ket{\psi_N^-} \\ \nonumber
&+\frac{\omega_{\mathrm{q}}}{2}\sum_{N=1,3,..}\frac{(2\alpha)^N}{(\mathcal{E}_{N+}+\omega\alpha^2+\omega_{\mathrm{q}}/2\braket{0-|0+})\sqrt{N!}} \ket{\psi_N^+}\bigg)
\end{align}
\textcolor{black}{Then, by assuming $\omega_{\mathrm{q}}\ll \omega$, we have }
\begin{align}
\textcolor{black}{\mathcal{E}_{N\pm}}&\textcolor{black}{=\omega_{\mathrm{r}}(N-\alpha^2)\pm\frac{\omega_{\mathrm{q}}}{2}\braket{N-|N+} }\\
&\textcolor{black}{\approx \omega_{\mathrm{r}}(N-\alpha^2)}
\end{align}
\textcolor{black}{And, we have}
\begin{align} \label{gg}
\ket{G}\approx & \frac{1}{\sqrt{\mathcal{N}}} \bigg(\ket{\psi_0^-}
+\frac{\omega_{\mathrm{q}}}{2}e^{-2\alpha^2}\big(\sum_{N=2,4,..}\frac{(2\alpha)^N}{\omega N \sqrt{N!}}\ket{\psi_N^-} \\ \nonumber
&+\sum_{N=1,3,..}\frac{(2\alpha)^N}{\omega N \sqrt{N!}}\ket{\psi_N^+}\big)\bigg)
\end{align}
\textcolor{black}{where 
\begin{equation}
\mathcal{N}=1+\frac{\omega_{\mathrm{q}}^2}{\omega_{\mathrm{r}}^2}e^{-4\alpha^2}\sum_{N=1}^{\infty} \frac{(4\alpha^2)^N}{N^2 N !}.
\end{equation} Also, we set $\frac{\omega_{\mathrm{q}}}{2}\braket{0-|0+} \approx 0$  }
\par
\textcolor{black}{Similarly, with regard to the first excited state, we can obtain }
\begin{align}
\ket{E}\approx & \frac{1}{\sqrt{\mathcal{N}}} \bigg[\ket{\psi_0^+}
-\frac{\omega_{\mathrm{q}}}{2}e^{-2\alpha^2}\bigg(\sum_{N=2,4,..}\frac{(2\alpha)^N}{\omega N \sqrt{N!}}\ket{\psi_N^+} \\ \nonumber
&+\sum_{N=1,3,..}\frac{(2\alpha)^N}{\omega N \sqrt{N!}}\ket{\psi_N^-}\bigg)\bigg]
\label{ee}
\end{align}
(Note that $\mathcal{N}$ in Eq.~\ref{gg} and Eq.~\ref{ee} are the same.)
The fidelity $F_{\mathrm{G}}=|\braket{\psi_0^-|G}|^2$ and $F_{\mathrm{E}}=|\braket{\psi_0^+|E}|^2$ are calculated as
\begin{align}
\textcolor{black}{F_{\mathrm{G}}=F_{\mathrm{E}}}&\textcolor{black}{=\frac{1}{\mathcal{N}}} \\
&\textcolor{black}{=\frac{1}{1+\frac{\omega_{\mathrm{q}}^2}{4\omega_{\mathrm{r}}^2}e^{-4\alpha^2}\sum_{N=1}^{\infty} \frac{(4\alpha^2)^N}{N^2 N !}}}
\end{align}
\textcolor{black}{Then, we define}
\begin{align}
\textcolor{black}{f \equiv \frac{\omega_{\mathrm{q}}^2}{4\omega_{\mathrm{r}}^2}\mathrm{exp}[-4\alpha^2]\sum_{N=1}^{\infty} \frac{(4\alpha^2)^N}{N^2 N !}}
\end{align}
\textcolor{black}{For $f \ll 1$, we have $F_{\mathrm{G}}=F_{\mathrm{E}}
\approx 1-f$, and so we can consider $f$ as an infidelity. }
\par
\begin{figure}[t]
\centering
\includegraphics[width=\columnwidth]{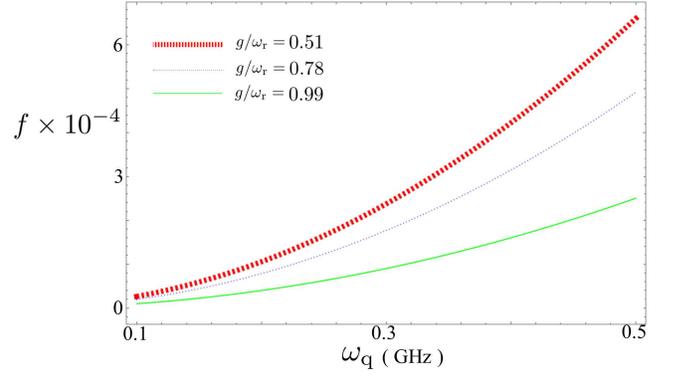}
\caption{\textcolor{black}{The infidelity $f$  versus qubit frequency $\omega_q$ for three regimes $g/\omega_{\mathrm{r}}=0.51$ (red dashed), $g/\omega_{\mathrm{r}}=0.78$ (blue dotted),  $g/\omega_{\mathrm{r}}=0.99$ (green lined).Here $\omega_{\mathrm{q}}$ varies from $2\pi \times 0.1 \mathrm{GHz}$ to $2\pi \times 0.5 \mathrm{GHz}$, where $\omega_{\mathrm{r}}=2\pi \times 6.336~ \mathrm{GHz} $. }}
\label{analy}
\end{figure}

\textcolor{black}{We plot $f$ for three regimes  $g/\omega_{\mathrm{r}}=0.51, 0.78, 0.99 $. Here, we fix $\omega_{\mathrm{r}}=2\pi \times 6.336~ \mathrm{GHz}  $ . From Fig. \ref{analy}, we can see that in these regimes the infidelity $f$ is sufficiently small.}
\vspace{5mm}

\begin{figure}[t]
\centering
\includegraphics[width=\columnwidth]{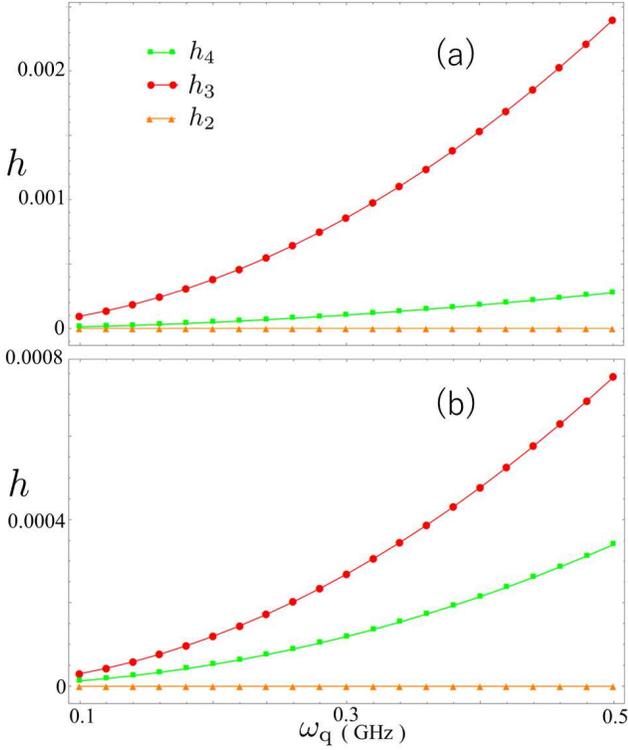}
\caption{\textcolor{black}{A measure of leakage to excited states $h_{2(3,4)}$   versus qubit frequency $\omega_q$. Again$\omega_{\mathrm{q}}$ varies from $2\pi \times 0.1 \mathrm{GHz}$ to $2\pi \times 0.5 \mathrm{GHz}$. We set (a) $g/\omega_{\mathrm{r}}=0.51$, (b) $g/\omega_{\mathrm{r}}=0.99$, where $\omega_{\mathrm{r}}=2\pi \times 6.336~ \mathrm{GHz} $.} }
\label{numer}
\end{figure}

\textcolor{black}{Also, we plot the numerically calculated
\textcolor{black}{\begin{equation} h_{j}=|\bra{\phi_{j}}\hat{\sigma}_z\ket{G}|^2, \end{equation}
$(j=2,3,4)$} in Fig. \ref{numer}  in the same regime where
$\ket{\phi_2}$,  $\ket{\phi_3}$, $\ket{\phi_4}$ are the second, third
and fourth excited states, respectively.
 This shows the
leakage  from $\ket{G}$  to unwanted states. }
\textcolor{black}{From Fig. \ref{analy} and Fig. \ref{numer}, we confirm
that, as long as  $f \ll 1$ is satisfied, transitions from $\ket{G}$
to the unwanted states such as $\ket{\phi_2}$,  $\ket{\phi_3}$, $\ket{\phi_4}$
are small, so that the two-level approximation should be valid in this regime.}

\section*{Appendix E:  Losses in the ultra-strongly coupled system}
\textcolor{black}{ So far we did not include a full analysis of the USC losses because we assumed that the time scale of such losses would be much longer than the readout time. For completeness,
here we present an short analysis of the influence of such losses.  Because including bath-induced transitions between all eigenstates
  in the full space is complex, here we restrict ourselves to the two-level approximation.  \textcolor{black}{\textcolor{black}{\textcolor{black}{We justify this approximation, in our relevant parameter regime, in the previous sections}}.}}
\par
\begin{figure}[t]
\centering
\vspace{2mm}
\includegraphics[width=\columnwidth]{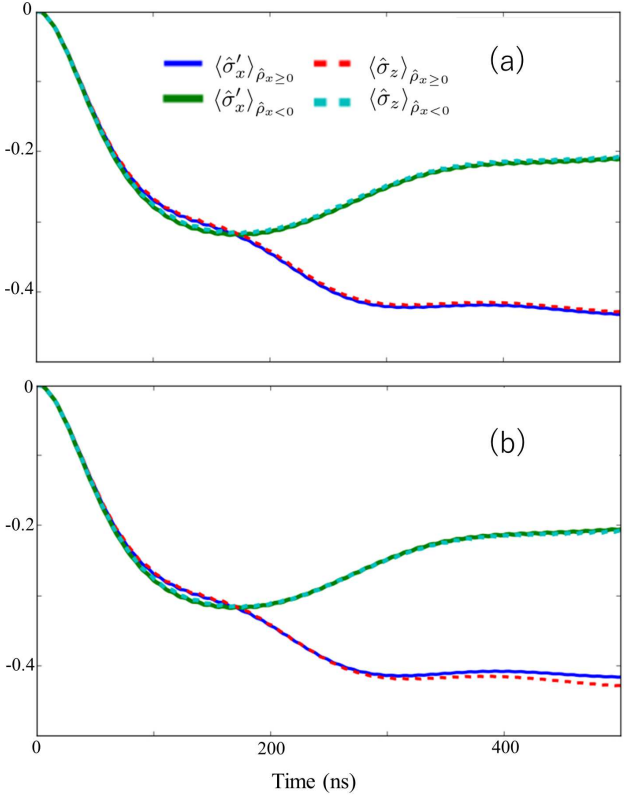}
\caption{\textcolor{black}{ The expectation values $\braket{\hat{\sigma}_z}$ (using the full Hamiltonian in Eq.~(\ref{total})) and $\braket{\hat{\sigma}_x^\prime}$ (using the approximate Hamiltonian in Eq.~(\ref{effhamiltonian})) (a) with and (b) without noise in the ultra strong system,
 after the coarse-graining measurements that projects the
 state into $\hat{\rho}_{x\geq0}$ or $\hat{\rho}_{x<0}$, depending on the measurement
 results. We set the coarse-graining value as $\sigma=5$. The noise rate $\gamma_1=\gamma_2=2 \pi \times 23.75 ~\mathrm{kHz}$. For the other parameters, we use the same as those in Fig. 2 in the main text.} }
\label{nouscdecay}
\end{figure}

\vspace{5mm}
\textcolor{black}{The interaction Hamiltonian between our ultra-strongly coupled system and environment is described as 
\begin{equation} 
\hat{H}_I=\hat{A}\hat{X}+\hat{B}\hat{\sigma}_z,
 \end{equation} 
 where $\hat{X}=\hat{a}+\hat{a}^\dag $ denotes the position operator and $\hat{A}$~($\hat{B}$) denotes the environmental operator coupled with the resonator (qubit). Also, we incorporate the effect of a dephasing bath classically modeled as 
 \begin{equation} \hat{H}_{\mathrm{dep}}=f(t) \hat{\sigma}_z,
 \end{equation} where $f(t)$ is a time-dependent random variable \textcolor{black}{\textcolor{black}{and the ensemble average of $f(t)$ is zero.}}
In this case, it is well know that the Born-Markov-Secular Lindblad master equation  can be written in the form \cite{ultra3} }
\begin{align}
\textcolor{black}{\frac{d \hat{\rho}}{dt} }&\textcolor{black}{=-i [\hat{H}_s, \hat{\rho}]} \\ \nonumber
&+\sum_{j,k>j} (\Gamma^{jk}_{\sigma_z}+\Gamma^{jk}_{X})\mathcal{D} [\ket{j}\bra{k}] (\hat{\rho}) \\ \nonumber
&+\sum_j \mathcal{D}[\Phi_j^{\mathrm{dep}}\ket{j}\bra{j}](\hat{\rho}) \\ \nonumber
&\textcolor{black}{+\sum_{j,k \neq j} \Gamma_{\mathrm{dep}}^{jk} \mathcal{D}[\ket{j}\bra{k}] (\hat{\rho}) } \\
\textcolor{black}{\mathcal{D}[\hat{A}](\hat{\rho})}&\textcolor{black}{=\frac{1}{2}(2\hat{A}\hat{\rho}\hat{A}^\dag-\hat{A}^\dag \hat{A}\hat{\rho}-\hat{\rho}\hat{A}^\dag \hat{A})} \\
\end{align}
\textcolor{black}{where $\hat{H}_s$ is the system Hamiltonian and
\begin{equation} \Gamma^{jk}_{\sigma_z}=\kappa_q
(\Delta_{jk})|\bra{j}\hat{\sigma}_z\ket{k}|^2,
\end{equation} and
\begin{equation}\Gamma^{jk}_{X}=\kappa_r(\Delta_{jk})|\bra{j}\hat{X}\ket{k}|^2.\end{equation} 
Here, $\ket{k}$ and $\ket{j}$ are the eigenstates of the system
Hamiltonian and $\kappa_q(\omega)$ and $\kappa_r(\omega)$ is the rate
corresponding to the noise spectra of the qubit and resonator,
respectively. Also, \begin{equation}\Phi_j
^{\mathrm{dep}}=\sqrt{\gamma_{\mathrm{dep}}(0)/2}\bra{j}\hat{\sigma}_x\ket{j}\end{equation}
\textcolor{black}{\textcolor{black}{and}
\begin{equation}\Gamma_{\mathrm{dep}}^{jk}=\gamma_{\mathrm{dep}}(\Delta_{jk})/2|\bra{j}\hat{\sigma}_x\ket{k}|^2,\end{equation}}
where $\gamma_{\mathrm{dep}}(\omega)$
\textcolor{black}{\textcolor{black}{ denotes the spectral density of the qubit dephasing at frequency $\omega$.}}
 Here, we ignore the term $ \Gamma_{\mathrm{dep}}^{jk} \mathcal{D}[\ket{j}\bra{k}] (\hat{\rho}) $ as this term is negligible when we operate at the ``sweet spot'' of the qubit. Owing to the two-level approximation, we consider only the lowest first two levels $\ket{G}$ and $\ket{E}$, and defining $\gamma_1=\Gamma^{10}_{\sigma_z}+\Gamma^{10}_{X} $, and $\gamma_2=(\Phi_0 ^{\mathrm{dep} })^2=(\Phi_1 ^{\mathrm{dep} })^2$, we obtain }
\begin{align}
\textcolor{black}{\frac{d \hat{\rho}}{dt}}&\textcolor{black}{=-i[\hat{H}^\prime, \hat{\rho}]} +\gamma_1~ \mathcal{D}[\ket{G}\bra{E}](\hat{\rho}) \\ \nonumber
&+\gamma_2~ \mathcal{D}[\ket{E}\bra{E}-\ket{G}\bra{G}](\hat{\rho})  \\
\textcolor{black}{\hat{H}^\prime}&\textcolor{black}{=\frac{\omega_{\mathrm{q}}}{2}\mathrm{exp}(-2 \alpha^2) (\ket{E}\bra{E}-\ket{G}\bra{G}) }
\end{align}
\textcolor{black}{where $\alpha=g/\omega_{\mathrm{r}}$.
\textcolor{black}{\textcolor{black}{In}} Fig.~\ref{nouscdecay}(a), we plot the expectation value of
$\hat{\sigma}_z$ and
$\hat{\sigma}_x^\prime=\ket{E}\bra{G}+\ket{G}\bra{E}$
without the noise in the ultra-strongly coupled system. The two-level
approximation shows an excellent agreement with the full Hamiltonian
model.}
\textcolor{black}{\textcolor{black}{Also, in Fig.~\ref{nouscdecay}(b), we plot $\hat{\sigma}_z$
and $\hat{\sigma}_x^\prime$ including the noise  in the
ultra-strongly-coupled system with parameters \cite{fqyou} that are realized in
recent experiments \cite{fqdecay}.
From these results, we can conclude that the noise in the ultra-strongly coupled system is almost negligible and does not have significance on the time scales in which we are interested.  } }
\par

\end{document}